\documentclass[12pt,journal,onecolumn]{IEEEtran}

\usepackage{amssymb,amsmath,amsfonts,bm,epsfig,graphicx,theorem,latexsym}
\usepackage{rotating,setspace,latexsym,epsf,color,subfigure}
\usepackage{cite,times,authblk,epstopdf}
\usepackage{multicol}%
\usepackage{verbatim}%
\usepackage{enumerate}
\usepackage{eucal}
\usepackage{caption}
\usepackage{longtable}

\newtheorem{theorem}{Theorem}

\newtheorem{corollary}{Corollary}
\newtheorem{Definition}{Definition}
\newtheorem{lemma}{Lemma}

\newenvironment{Proof}[1]{\medskip\par\noindent{\bf Proof:\,}\,#1}{{\mbox{\,$\blacksquare$}\medskip\par}}

\makeatletter

\allowdisplaybreaks[1]

\newcommand{\Xc}{\bold{X}_c}
\newcommand{\XcH}{\bold{X}_c^H}
\newcommand{\Xcn}{\bold{X}_c^n}

\newcommand{\Xctilde}{\tilde{\bold{X}}_c}
\newcommand{\Xcntilde}{\tilde{\bold{X}}_c^n}

\newcommand{\Xctwo}{\bold{X}_{c_2}}
\newcommand{\Xctdone}{\tilde{\bold{X}}_{c_1}}
\newcommand{\Xctdtwo}{\tilde{\bold{X}}_{c_2}}
\newcommand{\Xcntdone}{\tilde{\bold{X}}_{c_1}^n}
\newcommand{\Xcntdtwo}{\tilde{\bold{X}}_{c_2}^n}

\newcommand{\Xt}{\bold{X}_t}
\newcommand{\XtH}{\bold{X}_t^H}
\newcommand{\Xtn}{\bold{X}_t^n}

\newcommand{\Xtnset}{\mathcal{X}_t^n}

\newcommand{\Xttilde}{\tilde{\bold{X}}_t}
\newcommand{\Xtntilde}{\tilde{\bold{X}}_t^n}
\newcommand{\Xttdone}{\tilde{\bold{X}}_{t_1}}
\newcommand{\Xttdtwo}{\tilde{\bold{X}}_{t_2}}
\newcommand{\Xtntdone}{\tilde{\bold{X}}_{t_1}^n}
\newcommand{\Xtntdtwo}{\tilde{\bold{X}}_{t_2}^n}

\newcommand{\Xcdtdone}{{\tilde{\bold{X}}_{c_1}}'}
\newcommand{\Xcdtdtwo}{{\tilde{\bold{X}}_{c_2}}'}
\newcommand{\Xcdtdtwoone}{{\tilde{\bold{X}}_{c_{21}}}'}
\newcommand{\Xcdtdtwotwo}{{\tilde{\bold{X}}_{c_{22}}}'}
\newcommand{\Xcndtdone}{{\tilde{\bold{X}}_{c_1}}'^n}
\newcommand{\Xcndtdtwo}{{\tilde{\bold{X}}_{c_2}}'^n}
\newcommand{\Xcndtdtwoone}{\tilde{\bold{X}}_{c_{21}}'^n}
\newcommand{\Xcndtdtwotwo}{\tilde{\bold{X}}_{c_{22}}'^n}

\newcommand{\Yr}{\bold{Y}_r}
\newcommand{\Ye}{\bold{Y}_e}
\newcommand{\Yrn}{\bold{Y}_r^n}
\newcommand{\Yen}{\bold{Y}_e^n}
\newcommand{\Yrk}{Y_{r,k}}
\newcommand{\Yrkcon}{Y_{r,k}^{*}}
\newcommand{\Yrtilde}{\widetilde{\bold{Y}}_r}
\newcommand{\Yronetd}{\widetilde{Y}_{r_1}}
\newcommand{\YrtwoNtd}{\widetilde{\bold{Y}}_{r_2}^N}

\newcommand{\Yrhat}{\widehat{\bold{Y}}_r}
\newcommand{\Yrbardash}{\bar{\bar{\bold{Y}}}'_{r}}
\newcommand{\Yrhatdash}{\widehat{\bold{Y}}'_r}

\newcommand{\Yrbar}{\bar{\bar{\bold{Y}}}_r}
\newcommand{\Yrnbar}{\bold{\bar{Y}}_r^n}

\newcommand{\Zr}{\bold{Z}_r}
\newcommand{\Ze}{\bold{Z}_e}
\newcommand{\Zrn}{\bold{Z}_r^n}
\newcommand{\Zen}{\bold{Z}_e^n}
\newcommand{\Zttilde}{\tilde{\bold{Z}}_t}
\newcommand{\Zctilde}{\tilde{\bold{Z}}_c}
\newcommand{\Zctildetwo}{\tilde{\bold{Z}}_{c_2}}
\newcommand{\Ztntilde}{\tilde{\bold{Z}}_t^n}
\newcommand{\Zcntilde}{\tilde{\bold{Z}}_c^n}

\newcommand{\Zone}{\tilde{\bold{Z}}_1}
\newcommand{\Ztwo}{\tilde{\bold{Z}}_2}
\newcommand{\Zonen}{\tilde{\bold{Z}}_1^n}
\newcommand{\Ztwon}{\tilde{\bold{Z}}_2^n}
\newcommand{\Zrtilde}{\tilde{\bold{Z}}_r}
\newcommand{\Zrdtilde}{\tilde{\bold{Z}}'_r}
\newcommand{\Zetilde}{\tilde{\bold{Z}}_e}
\newcommand{\Zrntilde}{\tilde{\bold{Z}}_r^n}
\newcommand{\Zrndtilde}{\tilde{\bold{Z}}_r'^n}
\newcommand{\Zentilde}{\tilde{\bold{Z}}_e^n}

\newcommand{\Zedtilde}{\tilde{\bold{Z}}'_e}
\newcommand{\Zendtilde}{\tilde{\bold{Z}}_e'^n}

\newcommand{\ZrtwoN}{{\bold{Z}_r}_2^N}

\newcommand{\Ut}{\bold{U}_t}
\newcommand{\UtH}{\bold{U}_t^H}

\newcommand{\Vc}{\bold{V}_c}
\newcommand{\VcH}{\bold{V}_c^H}

\newcommand{\Uttwod}{{\bold{U}_t}_2^d}

\newcommand{\Utonel}{{\bold{U}_t}_1^l}
\newcommand{\Utlponed}{{\bold{U}_t}_{l+1}^d}

\newcommand{\Vconeg}{{\bold{V}_c}_1^g}

\newcommand{\Vctwol}{{\bold{V}_c}_2^l}
\newcommand{\Vcgponel}{{\bold{V}_c}_{g+1}^l}

\newcommand{\Uire}{U_{i,{{\rm{Re}}}}}
\newcommand{\Uiim}{U_{i,{{\rm{Im}}}}}
\newcommand{\Vire}{V_{i,{{\rm{Re}}}}}
\newcommand{\Viim}{V_{i,{{\rm{Im}}}}}
\newcommand{\Vkre}{V_{k,{{\rm{Re}}}}}
\newcommand{\Vkim}{V_{k,{{\rm{Im}}}}}

\newcommand{\Utwore}{U_{2,{{\rm{Re}}}}}
\newcommand{\Utwoim}{U_{2,{{\rm{Im}}}}}
\newcommand{\Vtwore}{V_{2,{{\rm{Re}}}}}
\newcommand{\Vtwoim}{V_{2,{{\rm{Im}}}}}

\newcommand{\Ulre}{U_{l,{{\rm{Re}}}}}
\newcommand{\Ulim}{U_{l,{{\rm{Im}}}}}
\newcommand{\Vlre}{V_{l,{{\rm{Re}}}}}
\newcommand{\Vlim}{V_{l,{{\rm{Im}}}}}

\newcommand{\Ht}{\bold{H}_t}
\newcommand{\Htbar}{\bar{\bold{H}}_t}
\newcommand{\HtH}{\bold{H}_t^H}

\newcommand{\Hc}{\bold{H}_c}
\newcommand{\HcH}{\bold{H}_c^H}

\newcommand{\Hcdash}{\bold{H}'_c}
\newcommand{\Hcbar}{\bar{\bold{H}}_c}
\newcommand{\Hctilde}{\widetilde{\bold{H}}_c}

\newcommand{\Hcone}{\bold{H}_{c_1}}
\newcommand{\Hctwo}{\bold{H}_{c_2}}
\newcommand{\HctwodH}{\bold{H}_{c_2}'^H}
\newcommand{\Hcdtwoone}{\bold{H}'_{c_{21}}}
\newcommand{\Hcdtwotwo}{\bold{H}_{c_{22}}}
\newcommand{\Hcom}{{\bold{H}}}
\newcommand{\Hcomh}{{\bold{H}}^H}

\newcommand{\bh}{\bold{h}}
\newcommand{\tbh}{\tilde{\bold{h}}}

\newcommand{\Gt}{\bold{G}_t}
\newcommand{\GtH}{\bold{G}_t^H}

\newcommand{\Gtpinv}{\bold{G}_t^{\dagger}}
\newcommand{\Gc}{\bold{G}_c}
\newcommand{\GcH}{\bold{G}_c^H}

\newcommand{\Gcom}{{\bold{G}}}

\newcommand{\Gcomp}{{\bold{G}}^{\sharp}}
\newcommand{\Gctilde}{\tilde{\bold{G}}_c}
\newcommand{\GctildeH}{\tilde{\bold{G}}_c^H}

\newcommand{\Gtone}{\bold{G}_{t_1}}
\newcommand{\Gttwo}{\bold{G}_{t_2}}
\newcommand{\Gcone}{\bold{G}_{c_1}}
\newcommand{\Gctwo}{\bold{G}_{c_2}}

\newcommand{\A}{\bold{A}}
\newcommand{\Atilde}{\widetilde{\bold{A}}}

\newcommand{\ba}{\bold{a}}
\newcommand{\tba}{\tilde{\bold{a}}}

\newcommand{\Pt}{\bold{P}_t}
\newcommand{\Ptbar}{\bar{\bold{P}}_t}
\newcommand{\Pta}{\bold{P}_{t,a}}
\newcommand{\Ptn}{\bold{P}_{t,n}}
\newcommand{\Ptone}{\bold{P}_{t,1}}
\newcommand{\Pttwo}{\bold{P}_{t,2}}
\newcommand{\PtH}{\bold{P}_t^H}
\newcommand{\Pc}{\bold{P}_c}
\newcommand{\Pcbar}{\bar{\bold{P}}_c}
\newcommand{\Pcone}{\bold{P}_{c,1}}
\newcommand{\Pctwo}{\bold{P}_{c,2}}
\newcommand{\PcH}{\bold{P}_c^H}
\newcommand{\PcI}{\bold{P}_{c,{\rm{I}}}}
\newcommand{\Pcn}{\bold{P}_{c,n}}
\newcommand{\bp}{\bold{p}}

\newcommand{\bz}{\bold{0}}
\newcommand{\bI}{\bold{I}}

\newcommand{\logdet}{\log{\rm{det}}}

\newcommand{\var}{{\rm{Var}}}
\newcommand{\E}{{\rm{E}}}
\newcommand{\Pro}{{\rm{Pr}}}

\newcommand{\Real}{{\rm{Re}}}
\newcommand{\Ima}{{\rm{Im}}}
\newcommand{\bb}{\bold{b}}
\newcommand{\bbH}{\bold{b}^H}
\newcommand{\dmin}{d_{\rm{min}}}

\newcommand{\Kt}{\bold{K}_t}
\newcommand{\Kc}{\bold{K}_c}

\title{Secure Degrees of Freedom for the MIMO Wire-tap Channel with a Multi-antenna Cooperative Jammer
\thanks{This paper was presented in part at the 2014 IEEE Information Theory Workshop, and the 2015 IEEE International Conference on Communications. This work was supported by NSF Grants CCF 09-64362, 13-19338 and CNS 13-14719.}}
\author{Mohamed Nafea}
\author{Aylin Yener}
\affil{\normalsize Wireless Communications and Networking Laboratory (WCAN)\\
Electrical Engineering Department\\
The Pennsylvania State University, University Park, PA 16802.\\
\em mnafea@psu.edu \quad\; yener@engr.psu.edu}

\begin{document}
\IEEEoverridecommandlockouts

\maketitle
\vspace{-2.3cm} 
\begin{center}\today\end{center}
 \vspace{.2cm}
\begin{abstract}
In this paper, a multiple antenna wire-tap channel in the presence of a multi-antenna cooperative jammer is studied. In particular, the secure degrees of freedom (s.d.o.f.) of this channel is established, with $N_t$ antennas at the transmitter, $N_r$ antennas at the legitimate receiver, and $N_e$ antennas at the eavesdropper, for all possible values of the number of antennas, $N_c$, at the cooperative jammer. In establishing the result, several different ranges of $N_c$ need to be considered separately. The lower and upper bounds for these ranges of $N_c$ are derived, and are shown to be tight. The achievability techniques developed rely on a variety of signaling, beamforming, and alignment techniques which vary according to the (relative) number of antennas at each terminal and whether the s.d.o.f. is integer valued. Specifically, it is shown that, whenever the s.d.o.f. is integer valued, Gaussian signaling for both transmission and cooperative jamming, linear precoding at the transmitter and the cooperative jammer, and linear processing at the legitimate receiver, are sufficient for achieving the s.d.o.f. of the channel. By contrast, when the s.d.o.f. is not an integer, the achievable schemes need to rely on structured signaling at the transmitter and the cooperative jammer, and joint signal space and signal scale alignment. The converse is established by combining an upper bound which allows for full cooperation between the transmitter and the cooperative jammer, with another upper bound which exploits the secrecy and reliability constraints. 
\end{abstract}

\section{Introduction}\label{Int}
Information theoretically secure message transmission in noisy communication channels was first considered in the seminal work by Wyner \cite{WTCWyner}. Reference \cite{CK} subsequently identified the secrecy capacity of a general discrete memoryless wire-tap channel. Reference \cite{leung1978gaussian} studied the Gaussian wire-tap channel and its secrecy capacity.  More recently, an extensive body of work was devoted to study a variety of network information theoretic models under secrecy constraint(s), see for example \cite{liu2008discrete,liu2009secrecy,ekrem2011secrecy2,khandani2013secrecy,
tekin2005secure,tekin2008general,liang2008multiple,bloch2013strong,gopala2008secrecy,
khisti2008secure,lai2008relay,oohama2007capacity,he2010cooperation,ekrem2011secrecy1,
liang2009compound,khisti2011interference,he2009k,he2014providing}. The secrecy capacity region for most of multi-terminal models remain open despite significant progress on bounds and associated insights. Recent work thus includes efforts that concentrate on characterizing the more tractable high signal-to-noise ratio (SNR) scaling behavior of secrecy capacity region for Gaussian multi-terminal models \cite{khisti2011interference,he2009k,he2014providing,xie2012secure,xie2013secure,xie2014secure}.

Among the multi-transmitter models studied, a recurrent theme in achievability is enlisting one or more terminals to transmit intentional interference with the specific goal of  diminishing the reception capability of the eavesdropper, known as  {\it{cooperative jamming}} \cite{Tekin2006}. For the Gaussian wire-tap channel, adding a cooperative jammer terminal transmitting Gaussian noise can improve the secrecy rate considerably \cite{tekin2008general}, albeit not the scaling of the secrecy capacity with power at high SNR. Recently, reference \cite{he2014providing} has shown that, for the Gaussian wire-tap channel, adding a cooperative jammer and utilizing structured codes for message transmission and cooperative jamming, provide an achievable secrecy rate scalable with power, i.e., a positive secure degrees of freedom (s.d.o.f.), an improvement from the zero degrees of freedom of the Gaussian wire-tap channel. More recently, reference \cite{xie2012secure} has proved that, for this channel, the s.d.o.f. $\frac{1}{2}$, achievable by codebooks constructed from integer lattices along with real interference alignment, is tight. References \cite{xie2013secure,xie2014secure} have subsequently identified the s.d.o.f. region for multi-terminal Gaussian wire-tap channel models. 

While the above development is for single-antenna terminals, multiple antennas have also been utilized to improve secrecy rates and s.d.o.f. for several channel models, see for example \cite{khisti2010secure,oggier2011secrecy,shafiee2009towards, liu2009secrecy,ekrem2011secrecy2,khandani2013secrecy,
khisti2011interference,liu2009note,he2014mimo,he2013mimo}. The secrecy capacity of the multi-antenna (MIMO) wire-tap channel, identified in\cite{khisti2010secure} scales with power only when the legitimate transmitter has an advantage over the eavesdropper in the number of antennas. It then follows naturally to utilize a cooperative jamming terminal to improve the secrecy rate and scaling for multi-antenna wire-tap channels as well which is the focus of this work.

In this paper, we study the multi-antenna wire-tap channel with a multi-antenna cooperative jammer. We characterize the high SNR scaling of the secrecy capacity, i.e., the s.d.o.f., of the channel with $N_c$ antennas at the cooperative jammer, $N_t$ antennas at transmitter, $N_r$ antennas at the receiver, and $N_e$ antennas at the eavesdropper. The achievability and converse techniques both are methodologically developed for ranges of the parameters, i.e., the number of antennas at each terminal. The upper and lower bounds for all parameter values are shown to match one another. The s.d.o.f. results in this paper match the achievability results derived in \cite{meGlobalSIP,meAllerton}, which are special cases for $\{N_t=N_r=1,N_c=N_e\}$, $\{N_t=N_r=N_e=N, N_c=2N\}$, $\{N_t=N_r=N_e=N, N_c=2N-1\}$, and real channel gains. The s.d.o.f. for the cases $\{N_t=N_r=N_e\}$ and $\{N_t=N_r\}$, for all possible values of $N_c$, were reported in \cite{meITW2014}, \cite{meICC2015}, respectively.

The proposed achievable schemes for different ranges of the values for $N_c$, $N_t$, $N_r$, and $N_e$ all involve linear precoding and linear receiver processing. The common goal to all these schemes is to perfectly align the cooperative jamming signals over the information signals observed at the eavesdropper while simultaneously enabling information and cooperative jamming signal separation at the legitimate receiver.  We show that whenever the s.d.o.f. of the channel is integer valued, Gaussian signaling both at the transmitter and the cooperative jammer suffices to achieve the s.d.o.f. By contrast, non-integer s.d.o.f. requires structured signaling along with joint signal space and signal scale alignment in the complex plane \cite{maddah2010degrees,kleinbock2002baker}. The necessity of structured signaling follows from the fact that fractional s.d.o.f. indicates sharing at least one spatial dimension between information and cooperative jamming signals at the receiver's signal space. In this case, sharing the same spatial dimension between Gaussian information and jamming signals, which have similar power scaling, does not provide positive degrees of freedom, and we need for structured signals that can be separated over this single dimension at high SNR. The tools that enable the signal scale alignment are available in the field of transcendental number theory \cite{kleinbock2002baker,Sprindzuk1,Sprindzuk2}, which we utilize. 

The paper is organized as follows. Section \ref{ChannelModel} introduces the channel model, and Section \ref{MainResult} provides the main results. For clarity of exposition, we first present the converse and achievability for the MIMO wire-tap channel with $N_t=N_r=N$ in Sections \ref{Conv_Proof} and \ref{AchSchemes}. Section \ref{Thm2_Proof} then extends the converse and achievability proofs for the case $N_t\neq N_r$. Section \ref{Discussion} discusses the results of this work and Section \ref{Con} concludes the paper.

Overall, this study determines the value in jointly utilizing signal scale and spatial interference alignment techniques for secrecy and quantifies the impact of a multi-antenna helper for the MIMO wire-tap channel by settling the question of the secrecy prelog for the $(N_t\times N_r\times N_e)$ MIMO wire-tap channel in the presence of an $N_c$-antenna cooperative jammer, for all possible values of $N_c$. In contrast with the single antenna case, where integer lattice codes and real interference alignment suffice to achieve the s.d.o.f. of the channel, in the MIMO setting, one needs to utilize a variety of signaling, beam-forming, and alignment techniques, in order to coordinate the transmitted and received signals for  different values of $N_t,N_r,N_e$, and $N_c$.

\section{Channel Model and Definitions}{\label{ChannelModel}}
First, we remark the notation we use throughout the paper: Small letters denote scalars and capital letters denote random variables. Vectors are denoted by bold small letters, while matrices and random vectors are denoted by bold capital letters\footnote{The distinction between matrices and random vectors is clear from the context.}. Sets are denoted using calligraphic fonts. All logarithms are taken to be base $2$. The set of integers $\left\{-Q,\cdots,Q\right\}$ is denoted by $(-Q,Q)_{\mathbb{Z}}$. $\bold{0}_{m\times n}$ denotes an $m\times n$ matrix of zeros, and $\bold{I}_{n}$ denotes an $n\times n$ identity matrix. For matrix $\A$, $\mathcal{N}(\A)$ denotes its null space, ${\rm{det}}(\A)$ denotes its determinant, and $||\A||$ denotes its {\it{induced}} norm. For vector $\bold{V}$, $||\bold{V}||$ denotes its Euclidean norm, and $\bold{V}_i^j$ denotes the $i$th to $j$th components in $\bold{V}$. We use $\bold{V}^n$ to denote the $n$-letter extension of the random vector $\bold{V}$, i.e., $\bold{V}^{n}=\left[\bold{V}(1)\;\cdots\bold{V}(n)\right]$. The operators $^T$, $^H$, and $^\dagger$ denote the transpose, Hermitian, and pseudo inverse operations. We use $\mathbb{R},\mathbb{C}$, $\mathbb{Q}$, and $\mathbb{Z}$, to denote the sets of real, complex, rational, and integer numbers, respectively. $\mathbb{Z}[j]$ denotes the set of Gaussian (complex) integers. A circularly symmetric Gaussian random vector with zero mean and covariance matrix $\bold{K}$ is denoted by $\mathcal{CN}(\bz,\bold{K})$. 

\begin{figure}
    \centering
	\includegraphics[width=13.5cm,height=7.5cm]{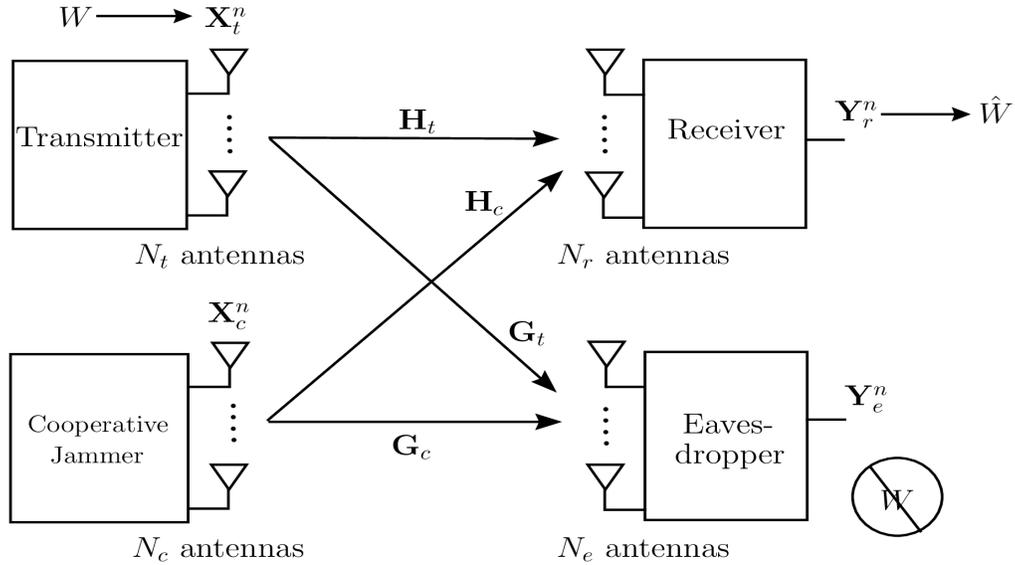}
	\caption{$(N_t\times N_r\times N_e)$ multiple antenna wire-tap channel with an $N_c$-antenna cooperative jammer.}
	\label{fig:sysmodel}
\end{figure}

As the channel model, we consider the MIMO wire-tap channel with an $N_t$-antenna transmitter, $N_r$-antenna receiver, $N_e$-antenna eavesdropper, and an $N_c$-antenna cooperative jammer as depicted in Fig. \ref{fig:sysmodel}. 
 The received signals at the receiver and eavesdropper, at the $n$th channel use, are given by
\begin{align}
\label{eq:Yr1}
\Yr(n)&=\Ht\Xt(n)+\Hc\Xc(n)+\Zr(n)\\
\label{eq:Ye1}
\Ye(n)&=\Gt\Xt(n)+\Gc\Xc(n)+\Ze(n),
\end{align}
where $\Xt(n)$ and $\Xc(n)$ are the transmitted signals from the transmitter and the cooperative jammer at the $n$th channel use. $\Ht\in\mathbb{C}^{N_r\times N_t}$, $\Hc\in \mathbb{C}^{N_r\times N_c}$ are the channel gain matrices from the transmitter and the cooperative jammer to the receiver, while $\Gt\in\mathbb{C}^{N_e\times N_t}$, $\Gc\in\mathbb{C}^{N_e\times N_c}$ are the channel gain matrices from the transmitter and the cooperative jammer to the eavesdropper. It is assumed that the channel gains are static, independently drawn from a {\it{complex-valued}} continuous distribution, and known at all terminals. $\Zr(n)$ and $\Ze(n)$ are the complex Gaussian noise at the receiver and eavesdropper at the $n$th channel use, where $\Zr(n)\sim\mathcal{CN}(\bold{0},\bold{I}_{N_r})$ and $\Ze(n)\sim\mathcal{CN}(\bold{0},\bold{I}_{N_e})$ for all $n$. $\Zr(n)$ is independent from $\Ze(n)$ and both are independent and identically distributed (i.i.d.) across the time index\footnote{Throughout the paper, we omit the index $n$ whenever possible.} $n$. The power constraints on the transmitted signals at the transmitter and the cooperative jammer are $\E\left\{\XtH\Xt\right\},\E\left\{\XcH\Xc\right\}\leq P$. 

The transmitter aims to send a message $W$ to the receiver, and keep it secret from the external eavesdropper. A stochastic encoder, which maps the message $W$ to the transmitted signal $\Xtn\in\Xtnset$, is used at the transmitter. The receiver uses its observation, $\Yrn\in\mathcal{Y}_r^n$, to obtain an estimate $\hat{W}$ of the transmitted message. Secrecy rate $R_s$ is achievable if for any $\epsilon>0$, there is a channel code $(2^{nR_s},n)$ satisfying\footnote{We consider weak secrecy throughout this paper.} 
\begin{align}
\label{eq:reliab_const}
&P_e={\rm{Pr}}\left\{\hat{W}\neq W\right\}\leq \epsilon,\\
\label{eq:sec_const}
&\frac{1}{n}H(W|\Yen)\geq \frac{1}{n}H(W)-\epsilon.
\end{align}
The secrecy capacity of a channel, $C_s$, is defined as the closure of all its achievable secrecy rates. For a channel with complex-valued coefficients, the achievable secure degrees of freedom (s.d.o.f.), for a given secrecy rate $R_s$, is defined as 
\begin{align}
\label{eq:sdof}
D_s=\underset{P\rightarrow\infty}\lim \frac{R_s}{\log P}.
\end{align}

The cooperative jammer transmits the signal $\Xcn\in\mathcal{X}_c^n$ in order to reduce the reception capability of the eavesdropper. However, this transmission affects the receiver as well, as interference.  
The jamming signal, $\Xcn$, does not carry any information. Additionally, there is no shared secret between the transmitter and the cooperative jammer.

\section{Main Result}\label{MainResult}
We first state the s.d.o.f. results for $N_t=N_r=N$. 

\begin{theorem}\label{Thm1}
The s.d.o.f. of the MIMO wire-tap channel with an $N_c$-antenna cooperative jammer, $N$ antennas at each of the transmitter and receiver, and $N_e$ antennas at the eavesdropper is given by 
\begin{align}
\label{eq:thm1}
D_s=\begin{cases}
[N+N_c-N_e]^+,\qquad \text{for}\;\; 0\leq N_c\leq N_e-\frac{\min\{N,N_e\}}{2}\\
N-\frac{\min\{N,N_e\}}{2},\qquad \text{for}\;\;  N_e-\frac{\min\{N,N_e\}}{2}< N_c \leq \max\{N,N_e\}\\
\frac{N+N_c-N_e}{2},\qquad \text{for}\;\; \max\{N,N_e\} < N_c\leq N+N_e.
\end{cases}
\end{align}
\end{theorem}

\begin{Proof}
The proof for Theorem \ref{Thm1} is provided in Sections \ref{Conv_Proof} and \ref{AchSchemes}.
\end{Proof}

Next, in Theorem \ref{Thm2} below, we generalize the result in Theorem \ref{Thm1} to $N_t\neq N_r$. 
\begin{theorem}\label{Thm2}
The s.d.o.f. of the MIMO wire-tap channel with an $N_c$-antenna cooperative jammer, $N_t$-antenna transmitter, $N_r$-antenna receiver, and $N_e$-antenna eavesdropper is given by
\begin{align}
\label{eq:thm2}
D_s=\begin{cases}
\min\left\{N_r,[N_c+N_t-N_e]^+\right\},\qquad \text{for}\;\; 0\leq N_c\leq N_1\\
\min\left\{N_t,N_r,\frac{N_r+\left[N_t-N_e\right]^+}{2}\right\},\qquad \text{for}\;\;  N_1< N_c \leq N_2\\
\min\left\{N_t,N_r,\frac{N_c+N_t-N_e}{2}\right\},\qquad \text{for}\;\; N_2 < N_c\leq N_3,
\end{cases}
\end{align}
where,
\begin{align*}
&N_1=\min\left\{N_e,\left[\frac{N_r}{2}+\frac{N_e-N_t}{2-1_{N_e>N_t}}\right]^+\right\},\quad 1_{N_e>N_t}=\begin{cases}
1,\qquad\text{if}\;\; N_e>N_t\\
0,\qquad\text{if}\;\;N_e\leq N_t
\end{cases}\\
&N_2=N_r+\left[N_e-N_t\right]^+,\quad N_3=\max\left\{N_2,2\min\left\{N_t,N_r\right\}+N_e-N_t\right\}.
\end{align*}
\end{theorem}
\begin{Proof}
The proof for Theorem \ref{Thm2} is provided in Section \ref{Thm2_Proof}.
\end{Proof}

{\bf{Remark 1}} Theorem \ref{Thm2} provides a complete characterization for the s.d.o.f. of the channel. The s.d.o.f. at $N_c=N_3$ is equal to $\min\{N_t,N_r\}$, which is equal to the d.o.f of the $(N_t\times N_r)$ point-to-point MIMO Gaussian channel. Thus, increasing the number of antenna at the cooperative jammer, $N_c$, over $N_3$ can not increase the s.d.o.f. over $\min\{N_t,N_r\}$. 

{\bf{Remark 2}} For $N_t\geq N_r+N_e$, the s.d.o.f. of the channel is equal to $N_r$ at $N_c=0$, i.e., the maximum s.d.o.f. of the channel is achieved without the help of the cooperative jammer. 

{\bf{Remark 3}} The converse proof for Theorem \ref{Thm2} involves combining two upper bounds for the s.d.o.f. derived for two different ranges of $N_c$. These two bounds are a straight forward generalization of those derived for the symmetric case in Theorem \ref{Thm1}. However, combining them is more tedious since more cases of the number of antennas at the different terminals should be handled carefully. Achievability for Theorem \ref{Thm2} utilizes similar techniques to those used for Theorem \ref{Thm1} as well, where handling more cases is required. For clarity of exposition, we derive the s.d.o.f. for the symmetric case first in order to present the main ideas, and then utilize these ideas and generalize the result to the asymmetric case of Theorem \ref{Thm2}. 

\begin{figure}
    \centering
	\includegraphics[width=13cm,height=7cm]{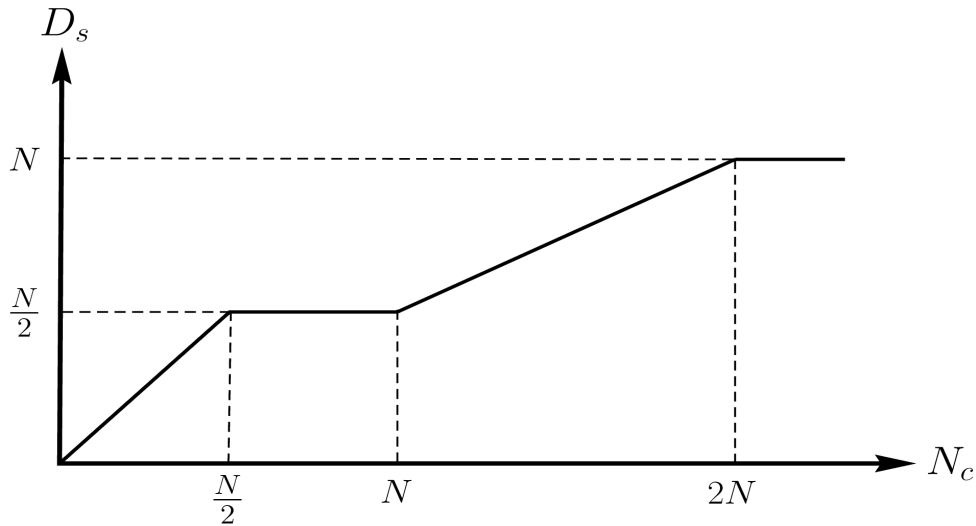}
	\caption{Secure degrees of freedom for a MIMO wire-tap channel, with $N$ antennas at each of its nodes, and a cooperative jammer with $N_c$ antennas, where $N_c$ varies from $0$ to $2N$.}
	\label{fig:sdof}
\end{figure}

For illustration purposes, the s.d.o.f. for $N_t=N_r=N_e=N$, and $N_c$ varies from $0$ to $2N$, is depicted in Fig. \ref{fig:sdof}. We provide the discussion of the results of this work in Section \ref{Discussion}.

\section{Converse for $N_t=N_r=N$}{\label{Conv_Proof}}
In Section \ref{Conv_Proof_1}, we derive the upper bound for the s.d.o.f. for $0\leq N_c\leq N_e$. In Section \ref{Conv_Proof_2}, we derive the upper bound for $\max\{N,N_e\}\leq N_c\leq N+N_e$. The two bounds are combined in Section \ref{Conv_Proof_3} to provide the desired upper bound in (\ref{eq:thm1}). 

\subsection{$0 \leq N_c\leq N_e$}{\label{Conv_Proof_1}}
Allow for full cooperation between the transmitter and the cooperative jammer. This cooperation can not decrease the s.d.o.f. of the channel, and yields a MIMO wire-tap channel with $N+N_c$-antenna transmitter, $N$-antenna receiver, and $N_e$-antenna eavesdropper. It has been shown in \cite{khisti2010secure} that, at high SNR, i.e., $P\rightarrow \infty$, the secrecy capacity of this channel, $C_s$, takes the asymptotic form 
\begin{align}
\label{eq:conv_1_1}
C_s(P)=\log {\rm{det}}\left(\bI_{N}+\frac{P}{p}\Hcom\Gcomp\Hcomh\right)+ o(\log P),
\end{align}
where $\underset{P\rightarrow \infty}\lim\frac{o(\log P)}{\log P}=0$, $\Hcom\in\mathbb{C}^{N\times (N+N_c)}$ and $\Gcom\in\mathbb{C}^{N_e\times (N+N_c)}$ are the channel gains from the combined transmitter to the receiver and eavesdropper, and $\Gcomp$ is the projection matrix onto the null space of $\Gcom$, $\mathcal{N}(\Gcom)$. $p={\rm{dim}}\left\{\mathcal{N}(\Hcom)^{\perp}\cap\mathcal{N}(\Gcom)\right\}$, where $\mathcal{N}(\Hcom)^{\perp}$ is the space orthogonal to the null space of $\Hcom$. Due to the randomly generated channel gains, if a vector $\bold{x}\in\mathcal{N}(\Gcom)$, then $\bold{x}\in\mathcal{N}(\Hcom)^{\perp}$ almost surely (a.s.), for all $0\leq N_c\leq N_e$. Thus, $p={\rm{dim}}(\mathcal{N}(\Gcom))=[N+N_c-N_e]^+$.

$\Hcom\Gcomp\Hcomh$ can be decomposed as
\begin{align}
\label{eq:conv_1_3}
\Hcom\Gcomp\Hcomh=\bold{\Psi}\begin{bmatrix}\bold{0}_{(N-p)\times (N-p)}&\bold{0}_{(N-p)\times p}\\\bold{0}_{p\times (N-p)}&\bold{\Omega}\end{bmatrix}\bold{\Psi}^H,
\end{align}
where $\bold{\Psi}\in \mathbb{C}^{N\times N}$ is a unitary matrix and $\bold{\Omega}\in\mathbb{C}^{p\times p}$ is a non-singular matrix \cite{khisti2010secure}. Let $\bold{\Psi}=\left[\bold{\Psi}_1\;\bold{\Psi}_2\right]$, where $\bold{\Psi}_1\in\mathbb{C}^{N\times (N-p)}$ and $\bold{\Psi}_2\in\mathbb{C}^{N\times p}$. Substituting (\ref{eq:conv_1_3}) in (\ref{eq:conv_1_1}) yields
\begin{align}
\label{eq:conv_1_4}
C_{s}(P)&= \logdet\left(\bI_{N}+\frac{P}{p}\bold{\Psi}_2 \bold{\Omega}\bold{\Psi}_2^H\right)+o(\log P)\\
\label{eq:conv_1_5}
&=\log {\rm{det}}\left(\bI_{p}+\frac{P}{p}\bold{\Omega}\bold{\Psi}_2^H\bold{\Psi}_2\right)+o(\log P)\\
\label{eq:conv_1_6}
&=\log P^p {\rm{det}}\left(\frac{1}{P}\bI_{p}+\frac{1}{p}\bold{\Omega}\right)+o(\log P)\\
\label{eq:conv_1_7}
&=p\log P + o(\log P),
\end{align}
where (\ref{eq:conv_1_5}) follows from Sylvester's determinant identity and (\ref{eq:conv_1_6}) follows from $\bold{\Psi}$ being unitary. 

The achievable secrecy rate of the original channel, $R_s$, is upper bounded by $C_s(P)$. Thus, the s.d.o.f. of the original channel, for $0\leq N_c\leq N_e$, is upper bounded as 
\begin{align}
\label{eq:conv_1_9}
D_s&=\underset{P\rightarrow\infty}\lim \frac{R_s}{\log P}\leq  \underset{P\rightarrow\infty}\lim \frac{p\log P+ o(\log P)}{\log P}\\
\label{eq:conv_1_11}
&=[N+N_c-N_e]^+.
\end{align}

\subsection{$\max\{N,N_e\}< N_c\leq N+N_e$}{\label{Conv_Proof_2}}
The upper bound we derive here is inspired by the converse of the single antenna Gaussian wire-tap channel with a single antenna cooperative jammer derived in \cite{xie2012secure}, though as we will see shortly, the vector channel extension resulting from multiple antennas does require care. Let $\phi_i$, for $i=1,2,\cdots,10$, denote constants which do not depend on the power $P$. 

The secrecy rate $R_s$ can be upper bounded as follows
\begin{align}
\label{eq:conv_2_1}
nR_s&= H(W)\\
\label{eq:conv_2_2}
&=H(W)-H(W|\Yen)+H(W|\Yen)-H(W|\Yrn)+H(W|\Yrn)\\
\label{eq:conv_2_3}
&\leq n\epsilon+ H(W|\Yen)-H(W|\Yrn,\Yen)+n\delta\\
\label{eq:conv_2_4}
&=I(W;\Yrn|\Yen)+n\phi_1\\
\label{eq:conv_2_5}
&=h(\Yrn|\Yen)-h(\Yrn|W,\Yen)+n\phi_1\\
\label{eq:conv_2_6}
&\leq h(\Yrn|\Yen)-h(\Yrn|W,\Yen,\Xtn,\Xcn)+n\phi_1\\
\label{eq:conv_2_7}
&=h(\Yrn,\Yen)-h(\Yen)-h(\Zrn)+n\phi_1,
\end{align}
where (\ref{eq:conv_2_3}) follows since $H(W)-H(W|\Yen)\leq n\epsilon$ by the secrecy constraint in (\ref{eq:sec_const}), $H(W|\Yrn)\leq n\delta$ by Fano's inequality, and $H(W|\Yrn)\geq H(W|\Yrn,\Yen)$ by the fact that conditioning does not increase entropy, (\ref{eq:conv_2_7}) follows since $\Zrn$ is independent from $\{W,\Yen,\Xtn,\Xcn\}$, and $\phi_1=\epsilon+\delta$.

Let $\Xttilde=\Xt+\Zttilde$ and $\Xctilde=\Xc+\Zctilde$, where $\Zttilde\sim\mathcal{CN}(\bz,\Kt)$ and $\Zctilde\sim\mathcal{CN}(\bold{0},\Kc)$. Note that $\Xttilde$ and $\Xctilde$ are noisy versions of the transmitted signals $\Xt$ and $\Xc$, respectively. $\Zttilde$ is independent from $\Zctilde$ and both are independent from $\{\Xt,\Xc,\Zr,\Ze\}$. $\Ztntilde$ and $\Zcntilde$ are i.i.d. sequences of the random vectors $\Zttilde$ and $\Zctilde$. In addition, let $\Zone=-\Ht\Zttilde-\Hc\Zctilde+\Zr$ and $\Ztwo=-\Gt\Zttilde-\Gc\Zctilde+\Ze$. Note that $\Zone\sim\mathcal{CN}(\bz,\bold{\Sigma}_{\Zone})$ and $\Ztwo\sim\mathcal{CN}(\bz,\bold{\Sigma}_{\Ztwo})$, where $\bold{\Sigma}_{\Zone}=\Ht\Kt\HtH+\Hc\Kc\HcH+\bI_{N}$ and $\bold{\Sigma}_{\Ztwo}=\Gt\Kt\GtH+\Gc\Kc\GcH+\bI_{N_e}$. $\Zonen$ and $\Ztwon$ are i.i.d. sequences of $\Zone$ and $\Ztwo$, since each of $\Zrn,\Zen,\Ztntilde,\Zcntilde$ is i.i.d. across time. The covariance matrices, $\Kt$ and $\Kc$, are chosen as $\Kt=\rho^2\bI_{N}$ and $\Kc=\rho^2\bI_{N_c}$, where $0<\rho\leq1/\max\left\{||\HcH||,\sqrt{||\GtH||^2+||\GcH||^2}\right\}$. This choice of $\Kt$ and $\Kc$ guarantees the finiteness $h(\Zttilde),h(\Zctilde),h(\Zone)$, and $h(\Ztwo)$ as shown in Appendix A. Starting from (\ref{eq:conv_2_7}), we have 
\begin{align}
\label{eq:conv_2_8}
n&R_s\leq h(\Yrn,\Yen)-h(\Yen)+n\phi_2\\
\label{eq:conv_2_9}
&=h(\Yrn,\Yen,\Xtntilde,\Xcntilde)-h(\Xtntilde,\Xcntilde|\Yrn,\Yen)-h(\Yen)+n\phi_2\\
\label{eq:conv_2_10}
&\leq h(\Xtntilde,\Xcntilde)+h(\Yrn,\Yen|\Xtntilde,\Xcntilde)-h(\Xtntilde,\Xcntilde|\Yrn,\Yen,\Xtn,\Xcn)-h(\Yen)+n\phi_2\\
\label{eq:conv_2_11}
&\leq h(\Xtntilde)+h(\Xcntilde)+h(\Yrn|\Xtntilde,\Xcntilde)+h(\Yen|\Xtntilde,\Xcntilde)-h(\Ztntilde,\Zcntilde)-h(\Yen)+n\phi_2\\
\label{eq:conv_2_12}
&=h(\Xtntilde)+h(\Xcntilde)+h(\Zonen|\Xtntilde,\Xcntilde)+h(\Ztwon|\Xtntilde,\Xcntilde)-h(\Yen)+n\phi_3\\
\label{eq:conv_2_13}
&\leq h(\Xtntilde)+h(\Xcntilde)+h(\Zonen)+h(\Ztwon)-h(\Yen)+n\phi_3\\
\label{eq:conv_2_14}
&=h(\Xtntilde)+h(\Xcntilde)-h(\Yen)+n\phi_4,
\end{align}
where (\ref{eq:conv_2_11}) follows since $\Ztntilde$ and $\Zcntilde$ are independent from $\{\Xtn,\Xcn,\Yrn,\Yen\}$, $\phi_2=\phi_1-h(\Zr)$, $\phi_3=\phi_2-h(\Zttilde)-h(\Zctilde)$, and $\phi_4=\phi_3+h(\Zone)+h(\Ztwo)$. We now consider the following two cases.

{\bf{Case 1: $N_e\leq N$}} 

We first lower bound $h(\Yen)$ in (\ref{eq:conv_2_14}) as follows. Using the infinite divisibility of Gaussian distribution, we can express a stochastically equivalent form of $\Ze$, denoted by $\bold{Z}'_e$, as  
\begin{align}
\label{eq:conv_2_15}
\bold{Z}'_e=\Gt\Zttilde+\Zetilde.
\end{align} 
where\footnote{The choice of $\Kt$ guarantees that $\bI_{N_e}-\Gt\Kt\GtH$ is a valid covariance matrix.} $\Zetilde\sim\mathcal{CN}(\bz,\bI_{N_e}-\Gt\Kt\GtH)$ is independent from $\{\Zttilde,\Zctilde,\Xt,\Xc,\Zr\}$. $\Zentilde$ is an i.i.d. sequence of the random vectors $\Zetilde$. Using (\ref{eq:conv_2_15}), a stochastically equivalent form of $\Yen$ is   
\begin{align}
\label{eq:conv_2_16}
{\bold{Y}_{e}'}^n=\Gt\Xtntilde+\Gc\Xcn+\Zentilde.
\end{align}

Let $\Xt=\left[X_{t,1}\cdots X_{t,N}\right]^T$, $\Zttilde=[\tilde{Z}_{t,1}\cdots\tilde{Z}_{t,N}]^T$, and $\Xttilde=[\Xttdone^T\;\Xttdtwo^T]^T$, where $\Xttdone=[\tilde{X}_{t,1}\cdots\tilde{X}_{t,{N_e}}]^T$, $\Xttdtwo=[\tilde{X}_{t,{N_e+1}}\cdots\tilde{X}_{t,N}]^T$, and $\tilde{X}_{t,k}=X_{t,k}+\tilde{Z}_{t,k}$, $k=1,2,\cdots,N$. In addition, let $\Gt=\left[\Gtone\;\Gttwo\right]$, where $\Gtone\in\mathbb{C}^{N_e\times N_e}$ and $\Gttwo\in\mathbb{C}^{N_e\times (N-N_e)}$. Using (\ref{eq:conv_2_16}), we have 
\begin{align}
\label{eq:conv_2_17}
h&(\Yen)=h({\bold{Y}_{e}'}^n)=h(\Gt\Xtntilde+\Gc\Xcn+\Zentilde)\\
\label{eq:conv_2_18}
&\geq h(\Gt\Xtntilde)=h(\Gtone\Xtntdone+\Gttwo\Xtntdtwo)\\
\label{eq:conv_2_19}
&\geq h(\Gtone\Xtntdone+\Gttwo\Xtntdtwo|\Xtntdtwo)=h(\Gtone\Xtntdone|\Xtntdtwo)\\
\label{eq:conv_2_20}
&=h(\Xtntdone|\Xtntdtwo)+n\log|\det(\Gtone)|.
\end{align}
where the inequality in (\ref{eq:conv_2_18}) follows since $\{\Gt\Xtntilde\}$ and $\{\Gc\Xcn+\Zentilde\}$ are independent, as for two independent random vectors $\bold{X}$ and $\bold{Y}$, we have $h(\bold{X}+\bold{Y})\geq h(\bold{X})$.

Substituting (\ref{eq:conv_2_20}) in (\ref{eq:conv_2_14}) results in
\begin{align}
\label{eq:conv_2_21}
nR_s&\leq h(\Xtntdone,\Xtntdtwo)+h(\Xcntilde)-h(\Xtntdone|\Xtntdtwo)-n\log|\det(\Gtone)|+n\phi_4\\
\label{eq:conv_2_22}
&=h(\Xtntdtwo)+h(\Xcntilde)+n\phi_5,
\end{align}
where $\phi_5=\phi_4-\log|\det(\Gtone)|$.

We now exploit the reliability constraint in (\ref{eq:reliab_const}) to derive another upper bound for $R_s$, which we combine with the bound in (\ref{eq:conv_2_22}) in order to obtain the desired bound for the s.d.o.f. when $N_e<N$ and $N\leq N_c\leq N+N_e$. The reliability constraint in (\ref{eq:reliab_const}) can be achieved only if \cite{cover2006elements}
\begin{align}
\label{eq:conv_2_23}
nR_s&\leq I(\Xtn;\Yrn)=h(\Yrn)-h(\Yrn|\Xtn)\\
\label{eq:conv_2_24}
&=h(\Yrn)-h(\Hc\Xcn+\Zrn).
\end{align}
Similar to (\ref{eq:conv_2_15}), a stochastically equivalent form of $\Zr$ is given by
\begin{align}
\label{eq:conv_2_25}
\Zr'=\Hc\Zctilde+\Zrtilde,
\end{align}
where\footnote{The choice of $\Kc$ guarantees that $\bI_N-\Hc\Kc\HcH$ is a valid covariance matrix.} $\Zrtilde\sim\mathcal{CN}(\bz,\bI_{N}-\Hc\Kc\HcH)$ is independent from $\{\Zttilde,\Zctilde,\Xt,\Xc,\Ze\}$. $\Zrntilde$ is an i.i.d. sequence of the random vectors $\Zrtilde$.

Let $\Xc=\left[X_{c,1}\cdots X_{c,N_c}\right]^T$, $\Zctilde=[\tilde{Z}_{c,1}\cdots\tilde{Z}_{c,N_c}]^T$, and $\Xctilde=[\Xctdone^T\;\Xctdtwo^T]^T$, where $\Xctdone=[\tilde{X}_{c,1}\;\dots\;\tilde{X}_{c,N}]^T$, $\Xctdtwo=[\tilde{X}_{c,N+1}\;\cdots\;\tilde{X}_{c,N_c}]^T$, and $\tilde{X}_{c,k}=X_{c,k}+\tilde{Z}_{c,k}$, $k=1,2,\cdots,N_c$. In addition, let $\Hc=\left[\Hcone\;\Hctwo\right]$, where $\Hcone\in\mathbb{C}^{N\times N}$ and $\Hctwo=\in\mathbb{C}^{N\times (N_c-N)}$. Using (\ref{eq:conv_2_25}), we have   
\begin{align}
\label{eq:conv_2_26}
h(&\Hc\Xcn+\Zrn)=h(\Hc\Xcn+{\bold{Z}_r'}^n)=h(\Hc\Xcntilde+\Zrntilde)\\
\label{eq:conv_2_27}
&\geq h(\Hc\Xcntilde)=h(\Hcone\Xcntdone+\Hctwo\Xcntdtwo)\\
\label{eq:conv_2_28}
&\geq  h(\Hcone\Xcntdone|\Xcntdtwo)\\
\label{eq:conv_2_29}
&=h(\Xcntdone|\Xcntdtwo)+n\log|\det(\Hcone)|.
\end{align}
 
Substituting (\ref{eq:conv_2_29}) in (\ref{eq:conv_2_24}) yields
\begin{align}
\label{eq:conv_2_30}
nR_s\leq h(\Yrn)-h(\Xcntdone|\Xcntdtwo)-n\log|\det(\Hcone)|.
\end{align}
Let $\Yr=\left[Y_{r,1}\;\cdots\;Y_{r,N}\right]^T$. Summing (\ref{eq:conv_2_22}) and (\ref{eq:conv_2_30}) results in 
\begin{align}
\label{eq:conv_2_31}
nR_s&\leq\frac{1}{2}\left\{h(\Yrn)+h(\Xtntdtwo)+h(\Xcntdtwo)\right\}+n\phi_6\\
\label{eq:conv_2_32}
&\leq\frac{1}{2}\sum_{i=1}^n\left\{\sum_{k=1}^N h(Y_{r,k}(i))+\sum_{k=N_e+1}^N h(\tilde{X}_{t,k}(i))+\sum_{k=N+1}^{N_c} h(\tilde{X}_{c,k}(i))\right\}+n\phi_6,
\end{align}
where $\phi_6=\frac{1}{2}\left(\phi_5-\log|\det(\Hcone)|\right)$.

In Appendix B, we show, for $i=1,\cdots,n$, $k=1,\cdots,N$, and $j=1,\cdots,N_c$, that 
\begin{align}
\label{eq:conv_2_33}
&h(Y_{r,k}(i))\leq \log 2\pi e+\log (1+h^2 P)\\
\label{eq:conv_2_34}
&h(\tilde{X}_{t,k}(i)), h(\tilde{X}_{c,j}(i))\leq \log 2\pi e+\log (\rho^2+P),
\end{align}
where $h^2=\underset{k}{\max}\;\left(||\bh_{t,k}^{r}||^2+||\bh_{c,k}^{r}||^2\right)$; $\bh_{t,k}^{r}$ and $\bh_{c,k}^{r}$ denote the transpose of the $k$th row vectors of $\Ht$ and $\Hc$, respectively. Using (\ref{eq:conv_2_32}), (\ref{eq:conv_2_33}), and (\ref{eq:conv_2_34}), we have 
\begin{align}
\label{eq:conv_2_35}
R_s\leq \frac{N}{2}\log (1+h^2 P)+\frac{N_c-N_e}{2}\log(\rho^2+P)+\phi_7,
\end{align}
where $\phi_7=\phi_6+\frac{N+N_c-N_e}{2}\log 2\pi e$. Using (\ref{eq:sdof}), we get
\begin{align}
\label{eq:conv_2_36}
D_s&\leq \underset{P\rightarrow\infty}\lim \frac{\frac{N}{2}\log (1+h^2 P)+\frac{N_c-N_e}{2}\log(\rho^2+P)+\phi_7}{\log P}\\
\label{eq:conv_2_37}
&=\frac{N+N_c-N_e}{2}.
\end{align}
Thus, the s.d.o.f. for $N_e\leq N$ and $N\leq N_c\leq N+N_e$, is upper bounded by $\frac{N+N_c-N_e}{2}$.

{\bf{Case 2: $N_e>N$}} \\
Another stochastically equivalent form of $\Ze$ is 
\begin{align}
\label{eq:conv_2_38}
\bold{Z}''_e=\Gt\Zttilde+\Gc\Zctilde+\Zedtilde.
\end{align} 
where\footnote{The choice of $\Kt$ and $\Kc$ guarantees that $\bI_{N_e}-\Gt\Kt\GtH-\Gc\Kc\GcH$ is a valid covariance matrix.} $\Zedtilde\sim\mathcal{CN}(\bz,\bI_{N_e}-\Gt\Kt\GtH-\Gc\Kc\GcH)$ is independent from $\{\Zttilde,\Zctilde,\Xt,\Xc,\Zr\}$. $\Zendtilde$ is an i.i.d. sequence of the random vectors $\Zedtilde$. Using (\ref{eq:conv_2_38}), another stochastically equivalent form of $\Yen$ is given by    
\begin{align}
\label{eq:conv_2_39}
{{\bold{Y}}''_{e}}^n=\Gt\Xttilde+\Gc\Xcntilde+\Zendtilde.
\end{align}

Let us rewrite $\Xctilde$ and $\Hc$ as follows. $\Xctilde=[\Xcdtdone^T\;\Xcdtdtwo^T]^T$, where $\Xcdtdone=[\tilde{X}_{c,1}\cdots\tilde{X}_{c,{N_e-N}}]^T$, $\Xcdtdtwo=[\Xcdtdtwoone^T\;\Xcdtdtwotwo^T]^T$, $\Xcdtdtwoone=[\tilde{X}_{c,{N_e-N+1}}\cdots\tilde{X}_{c,{N_e}}]^T$, and $\Xcdtdtwotwo=[\tilde{X}_{c,{N_e+1}}\cdots\tilde{X}_{c,{N_c}}]^T$. $\Hc=[\Hcone'\;\Hctwo']$, where $\Hcone'\in\mathbb{C}^{N\times {(N_e-N)}}$, $\Hctwo'=[\Hcdtwoone\;\Hcdtwotwo]$, $\Hcdtwoone\in\mathbb{C}^{N\times N}$, and $\Hcdtwotwo\in\mathbb{C}^{N\times (N_c-N_e)}$. Let $\Gc=\left[\Gcone\;\Gctwo\right]$, where $\Gcone\in\mathbb{C}^{N_e\times(N_e-N)}$ and $\Gctwo\in\mathbb{C}^{N_e\times (N+N_c-N_e)}$. Using (\ref{eq:conv_2_39}), we have 
\begin{align}
\label{eq:conv_2_40}
h(\Yen)&=h({{\bold{Y}}''_{e}}^n)=h([\Gt\;\Gcone]\begin{bmatrix}\Xtntilde\\\Xcndtdone\\\end{bmatrix}+\Gctwo\Xcndtdtwo+\Zendtilde)\\
\label{eq:conv_2_41}
&\geq h(\Xtntilde,\Xcndtdone|\Xcndtdtwo)+n\log|\det[\Gt\;\Gcone]|\\
\label{eq:conv_2_42}
&\geq h(\Xtntilde)+h(\Xcndtdone|\Xcndtdtwo)+n\log|\det[\Gt\;\Gcone]|,
\end{align} 
where (\ref{eq:conv_2_42}) follows since $\Xtntilde$ and $\Xcndtdtwo$ are independent. Substituting (\ref{eq:conv_2_42}) in (\ref{eq:conv_2_14}) gives
\begin{align}
\label{eq:conv_2_43}
nR_s\leq h(\Xcndtdtwo)+n\phi_8,
\end{align}
where $\phi_8=\phi_4-\log|\det[\Gt\;\Gcone]|$.

In order to obtain another upper bound for $R_s$, which we combine with (\ref{eq:conv_2_43}) to obtain the desired bound for $N_e>N$ and $N_e\leq N_c\leq N+N_e$, we proceed as follows. Consider a modified channel where the first $N_e-N$ antennas at the cooperative jammer are removed, i.e., the cooperative jammer uses only the last $N+N_c-N_e$ out of its $N_c$ antennas. The transmitted signals in the modified channel are $\Xtn$ and $\bold{X}_{c_2}'^{n}$, and hence, the legitimate receiver receives
\begin{align}
\label{eq:modified_channel}
\Yrnbar=\Ht\Xtn+\Hctwo'\bold{X}_{c_2}'^{n}+\Zrn. 
\end{align}
Since the cooperative jamming signal is additive interference for the legitimate receiver, the reliable communication rate of this modified channel, $\bar{R}$, is an upper bound for that of the original channel, $R$. Since $R_s$ satisfies the reliability and secrecy constraints in (\ref{eq:reliab_const}) and (\ref{eq:sec_const}), we have that  
\begin{align}
\label{eq:conv_2_44}
nR_s&\leq nR\leq n\bar{R}\leq I(\Xtn;\Yrnbar)= h(\Yrnbar)-h(\Hctwo'\bold{X}_{c_2}'^{n}+\Zrn).
\end{align}
Let $\Zctildetwo=[\tilde{Z}_{c,{N_e-N+1}}\cdots \tilde{Z}_{c,{N_c}}]^T\sim\mathcal{CN}(\bz,\Kc')$, where $\Kc'=\rho^2\bI_{N+N_c-N_e}$. Another stochastically equivalent form of $\Zr$ is $\Zr''=\Hctwo'\Zctildetwo+\Zrdtilde$, where\footnote{The choice of $\Kc$ guarantees that $\bI_N-\Hctwo'\Kc'\HctwodH$ is a valid covariance matrix.} $\Zrdtilde\sim\mathcal{CN}(\bz,\bI_{N}-\Hctwo'\Kc'\HctwodH)$ is independent from $\{\Zttilde,\Zctilde,\Xt,\Xc,\Ze\}$, and $\Zrndtilde$ is an i.i.d. sequence of $\Zrdtilde$. Thus, using (\ref{eq:conv_2_44}), we have 
\begin{align}
\label{eq:conv_2_44_1}
nR_s&\leq h(\Yrnbar)-h(\Hctwo'\Xcndtdtwo+\Zrndtilde)\leq h(\Yrnbar)-h(\Hctwo'\Xcndtdtwo)\\
\label{eq:conv_2_45}
&\leq h(\Yrnbar)-h(\Xcndtdtwoone|\Xcndtdtwotwo)-n\log|\det(\bold{H}_{c_{21}}')|.
\end{align}

Let $\bold{\bar{Y}}_r=[\bar{Y}_{r,1}\cdots \bar{Y}_{r,N}]^T$. Summing (\ref{eq:conv_2_43}) and (\ref{eq:conv_2_45}) yields
\begin{align}
\label{eq:conv_2_46}
nR_s&\leq \frac{1}{2}\left\{h(\bold{\bar{Y}}_r^n)+h(\Xcndtdtwotwo)\right\}+n\phi_9\\
\label{eq:conv_2_47}
&\leq \frac{1}{2}\sum_{i=1}^n\left\{\sum_{k=1}^N h(\bar{Y}_{r,k}(i))+\sum_{k=N_e+1}^{N_c} h(\tilde{X}_{c,k}(i))\right\}+n\phi_9,
\end{align} 
where $\phi_9=\frac{1}{2}\{\phi_8-\log|\det(\Hcdtwoone)|\}$. In Appendix. B, we also show that
\begin{align}
\label{eq:conv_2_48}
h(\bar{Y}_{r,k}(i))\leq \log 2\pi e+\log (1+\bar{h}^2 P),
\end{align}
where $\bar{h}^2=\underset{k}{\max}\;\left(||\bh_{t,k}^{r}||^2+||\bh_{c,k}'^{r}||^2\right)$; $\bh_{c,k}'^{r}$ denotes the transpose of the $k$th row vector of $\Hctwo'$.

Similar to case $1$, using (\ref{eq:conv_2_47}), (\ref{eq:conv_2_48}), and  (\ref{eq:conv_2_34}), the secrecy rate is bounded as 
\begin{align}
\label{eq:conv_2_49}
R_s\leq \frac{N}{2}\log(1+\bar{h}^2 P)+\frac{N_c-N_e}{2}\log(\rho^2+P)+n\phi_{10},
\end{align}
where $\phi_{10}=\phi_9+\frac{N+N_c-N_e}{2}\log 2\pi e$. Thus, the s.d.o.f., for $N_e>N$ and $N_e\leq N_c\leq N+N_e$, is upper bounded as 
\begin{align}
\label{eq:conv_2_50}
D_s\leq \frac{N+N_c-N_e}{2}.
\end{align} 

\subsection{Obtaining the Upper Bound} {\label{Conv_Proof_3}}
For $N_e\leq N$, the upper bound for the s.d.o.f. derived in Section \ref{Conv_Proof_1} is equal to $N+N_c-N_e$, for all $0\leq N_c\leq N_e$.  while the upper bound derived in Section \ref{Conv_Proof_2}, at $N_c=N$, is equal to $N-\frac{N_e}{2}$, c.f. equations (\ref{eq:conv_1_11}) and (\ref{eq:conv_2_37}). As the former upper bound is greater than the latter for all $\frac{N_e}{2}<N_c\leq N$, the s.d.o.f. is upper bounded by $N-\frac{N_e}{2}$ for all $\frac{N_e}{2}<N_c\leq N$. Combining these statements, we have the following upper bound for the s.d.o.f. for $N_e\leq N$:
\begin{align}
\label{eq:conv_3_1}
D_s\leq \begin{cases}
N+N_c-N_e,\qquad \text{for}\;\; 0\leq N_c\leq \frac{N_e}{2}\\
N-\frac{N_e}{2},\qquad \text{for}\;\;  \frac{N_e}{2}< N_c \leq N\\
\frac{N+N_c-N_e}{2},\qquad \text{for}\;\; N< N_c\leq N+N_e.
\end{cases}
\end{align}

Similarly, when $N_e>N$ and for all $N_e-\frac{N}{2}<N_c\leq N_e$, the upper bound derived for $0\leq N_c\leq N_e$ in Section \ref{Conv_Proof_1} is greater than the upper bound derived in Section \ref{Conv_Proof_2} at $N_c=N_e$. Thus, the s.d.o.f. for $N_e-\frac{N}{2}<N_c\leq N_e$ is upper bounded by $\frac{N}{2}$. In addition, the upper bound in (\ref{eq:conv_1_11}) is equal to zero for all $0\leq N_c\leq N_e-N$. Thus, the s.d.o.f. for $N_e>N$ is upper bounded as: 
\begin{align}
\label{eq:conv_3_2}
D_s\leq \begin{cases}
0,\qquad\qquad\qquad \;\;\text{for}\;\; 0\leq N_c\leq N_e-N\\
N+N_c-N_e,\qquad \text{for}\;  N_e-N<N_c\leq N_e-\frac{N}{2}\\
\frac{N}{2},\qquad\qquad\qquad\; \text{for}\;\;N_e-\frac{N}{2}< N_c \leq N_e\\
\frac{N+N_c-N_e}{2},\qquad\qquad \text{for}\;\; N_e< N_c\leq N+N_e.
\end{cases}
\end{align}

By combining the bounds for $N_e\leq N$ in (\ref{eq:conv_3_1}) and for $N_e>N$ in (\ref{eq:conv_3_2}), we obtain the upper bound for the s.d.o.f. in (\ref{eq:thm1}). In the next Section, we will show the achievability of (\ref{eq:thm1}).

\section{Achievablility for $N_t=N_r=N$}\label{AchSchemes}
In this section, we provide the achievability proof for Theorem \ref{Thm1} by showing the achievability of (\ref{eq:conv_3_1}) when $N_e\leq N$, and the achievability of (\ref{eq:conv_3_2}) when $N_e>N$. For both $N_e\leq N$ and $N_e>N$, we divide the range of the number of antennas at the cooperative jammer, $N_c$, into five ranges and propose an achievable scheme for each range. For all the achievable schemes in this section, we have the $n$-letter signals, $\Xtn$ and $\Xcn$, as i.i.d. sequences. Since $\Xcn$ is independent from $\Xtn$, and each of them is i.i.d. across time, we have in effect a memoryless wire-tap channel and the secrecy rate 
\begin{align}
\label{eq:AchSecRate}
R_s=[I(\Xt;\Yr)-I(\Xt;\Ye)]^{+},
\end{align}   
is achievable by {\it{stochastic encoding}} at the transmitter \cite{CK}.

The transmitted signals at the transmitter and the cooperative jammer, for each of the following schemes, are 
\begin{align}
\label{eq:Xt1_Xc1}
\Xt&=\Pt\Ut,\quad \Xc=\Pc\Vc,
\end{align}
where $\Ut=\left[U_1\cdots U_d\right]^T$ and $\Vc=\left[V_1\cdots V_l\right]^T$ are the information and cooperative jamming streams, respectively. $\Pt=\left[\bp_{t,1}\;\cdots\bp_{t,d}\right]\in\mathbb{C}^{N\times d}$ and $\Pc=\left[\bp_{c,1}\cdots\bp_{c,l}\right]\in\mathbb{C}^{N_c\times l}$ are the precoding matrices at the transmitter and the cooperative jammer.

Signaling, precoding, and decoding techniques utilized in this proof vary according to the relative number of antennas at the different terminals and whether the s.d.o.f. of the channel is integer valued or not an integer. In particular, we show that Gaussian signaling both for transmission and cooperative jamming is sufficient to achieve the integer valued s.d.o.f., while achieving non-integer s.d.o.f. requires structured signaling and cooperative jamming along with a combination of linear receiver processing, and the complex field equivalent of real interference alignment \cite{maddah2010degrees,kleinbock2002baker}. Additionally, the linear precoding at the transmitter and the cooperative jammer depends on whether $N_e$ is equal to, smaller than, or larger than $N$, and whether the number of antennas at the cooperative jammer, $N_c$, results in a s.d.o.f. for the channel that is before, after, or at the flat s.d.o.f. range in the s.d.o.f. plot versus $N_c$. This leads to an achievability proof that involves $10$ distinct achievable schemes, which differ from each other in the type of signals used (Gaussian or structured), and/or precoding at the transmitter and cooperative jammer, and/or decoding at the legitimate receiver.   

In order to extend real interference alignment to complex channels, we need to utilize different results than those used for real channels. For real channels, to analyze the decoder performance, reference \cite{motahari2009real2} proposed utilizing the convergence part of Khintchine-Groshev theorem in the field of Diophantine approximation \cite{schmidt1980diophantine}, which deals with the approximation of real numbers with rational numbers. For complex channels, transforming the channel into a real channel with twice the dimensions, as is usually the convention, is not sufficient here, since real interference alignment relies on the linear independence over rational numbers of the channel gains, which does not continue to hold after such channel transformation. Luckily, we can utilize a result in the field of classification of transcendental complex numbers, which provides a bound on the absolute value of a complex algebraic number with rational coefficients in terms of its height, i.e., the maximum coefficient \cite{Sprindzuk1,Sprindzuk2,kleinbock2002baker}. For complex channel coefficients, this result ends up playing the same role of the Khintchine-Groshev theorem for real coefficients. 

Before continuing with the achievability proof for the different cases, we state the following lemma, which is utilized to show the linear independence between the directions of the received streams at the legitimate receiver.  
\begin{lemma}\label{lemma1}
Consider two matrices $\bold{E}_1\in\mathbb{C}^{N\times K}$ and $\bold{E}_2\in\mathbb{C}^{K\times M}$, where $N,M<K$. If the matrix $\bold{E}_2$ is full column rank and the matrix $\bold{E}_1$ has all of its entries independently and randomly drawn according to a continuous distribution, then ${\rm{rank}}(\bold{E}_1\bold{E}_2)=\min(N,M)$ a.s. 
\end{lemma}
\begin{Proof}
The proof of Lemma \ref{lemma1} is given in Appendix C.
\end{Proof}

\subsection{Case 1: $N_e\leq N$ and $0\leq N_c\leq \frac{N_e}{2}$}\label{AchScheme1}
The s.d.o.f. for this case is equal to $N+N_c-N_e$, i.e., integer valued, for which we utilize Gaussian signaling and cooperative jamming. Since $N_e\leq N$, the transmitter exploits this advantage by sending a part of its signal invisible to the eavesdropper. There is no need for linear precoding at the cooperative jammer for this case. Increasing the number of the cooperative jammer antennas, $N_c$, increases the s.d.o.f. of the channel. 
 
The transmitted signals, $\Xt$ and $\Xc$, are given by (\ref{eq:Xt1_Xc1}) with $d=N+N_c-N_e$, $l=N_c$, $\Ut\sim \mathcal{CN}(\bold{0},\bar{P}\bold{I}_d)$, $\Vc \sim\mathcal{CN}(\bold{0},\bar{P}\bold{I}_l)$, $\Pc=\bI_{l}$, and
\begin{align}
\label{eq:Pt}
\Pt=\left[\Pta\;\;\Ptn\right]\in\mathbb{C}^{N\times d},
\end{align}
where $\Pta=\Gtpinv\Gc$ in order to align the information streams over the cooperative jamming streams at the eavesdropper, and the $N-N_e$ columns of $\Ptn$ are chosen to span $\mathcal{N}(\Gt)$. $\bar{P}=\frac{1}{\alpha} P$, in accordance with the power constraints on the transmitted signals at the transmitter and the cooperative jammer, where $\alpha=\max\left\{l,\sum_{i=1}^d ||\bp_{t,i}||^2\right\}$ is a constant which does not depend on the power $P$. 

Since $N_c\leq \frac{N_e}{2}$, the total number of superposed received streams at the receiver, $2N_c+N-N_e$, is less than or equal to the number of its available spatial dimensions, $N$. Thus, the receiver can decode all the information and cooperative jamming streams at high SNR. Using (\ref{eq:Yr1}), (\ref{eq:Ye1}), and (\ref{eq:Xt1_Xc1}), the received signals at the receiver and the eavesdropper are 
\begin{align}
\label{eq:Yr3_2}
\Yr&=\begin{bmatrix}\Ht\Pt&\Hc\\\end{bmatrix}\begin{bmatrix}\Ut\\\Vc\\\end{bmatrix}+\Zr,\\
\label{eq:Ye3_2}
\Ye&=\begin{bmatrix}\Gt\Gtpinv\Gc&\bz_{N_e\times (N-N_e)}\\\end{bmatrix}\begin{bmatrix}\Utonel\\\Utlponed\\\end{bmatrix}+\Gc\Vc+\Ze\\
\label{eq:Ye3_3}
&=\Gc(\Utonel+\Vc)+\Ze.
\end{align}                                                                                                                                                                                                                                                                                                                                                                                                                                                                                                                                                                                                                                                                                                                                                                                                                                                                                                                                                                                                                                                                                                                                                                                                                                                                                                                                                                                                                                                                                                                                                                                                                                                                                                                                                                                                                                                                                                                                                                                                                                                                                                                                                                                                                                                                                                                                                                                                                                                                                                                                                                                                                                                                                                                                                                                                                                                                                                                                                                                                                                                                                                                                                                                                                                                                                                                                                                                                                                                                                   
We lower bound the secrecy rate in (\ref{eq:AchSecRate}) as follows. First, in order to compute $I(\Xt;\Yr)$, we show that the matrix $\left[\Ht\Pt\;\;\Hc\right]\in\mathbb{C}^{N\times (d+l)}$ in (\ref{eq:Yr3_2}) is full column-rank a.s. 

The columns of $\Pta=\Gtpinv\Gc$ are linearly independent a.s. due to the randomly generated channel gains, and the $N-N_e$ columns of $\Ptn$ are linearly independent as well, since they span an $N-N_e$-dimensional subspace. In addition, each of the columns of $\Pta$ is linearly independent from the columns of $\Ptn$ a.s. since $\Gt\Pta=\Gc$, and hence $\Gt\bp_{t_i}\neq \bz$ for all $i=1,2,\cdots,l$. Thus $\Pt=[\Pta\;\Ptn]$ is full column rank a.s. The matrix $\left[\Ht\Pt\;\;\Hc\right]$ can be written as 
\begin{align}
\label{eq:Ach_1_1}
\begin{bmatrix}\Ht\Pt&\Hc\\\end{bmatrix}=\begin{bmatrix}\Ht&\Hc\\\end{bmatrix}\begin{bmatrix}\Pt&\bz_{N\times l}\\\bz_{l\times d}&\bI_{l}\\\end{bmatrix}.
\end{align}
The matrix $\left[\Ht\;\;\Hc\right]$ has all of its entries independently and randomly drawn according to a continuous distribution, while the second matrix on the right hand side (RHS) of (\ref{eq:Ach_1_1}) is full column rank a.s. By applying Lemma \ref{lemma1} to (\ref{eq:Ach_1_1}), we have that the matrix $\left[\Ht\Pt\;\;\Hc\right]$ is full column rank a.s. Thus, using (\ref{eq:Yr3_2}), we obtain the lower bound
\begin{align}
\label{eq:Ach_1_2}
I(\Xt;\Yr)\geq d\log P+o(\log P).
\end{align}  

Next, using (\ref{eq:Ye3_3}), we upper bound $I(\Xt;\Ye)$ as follows:
\begin{align}
\label{eq:Ach_1_3}
I(\Xt;\Ye)&=h(\Ye)-h(\Ye|\Xt)\\
\label{eq:Ach_1_4}
&=h(\Gc(\Utonel+\Vc)+\Ze)-h(\Gc\Vc+\Ze)\\
\label{eq:Ach_1_5}
&=\log\frac{{\rm{det}}(\bI_{N_e}+2\bar{P}\Gc\GcH)}{{\rm{det}}(\bI_{N_e}+\bar{P}\Gc\GcH)}\\
\label{eq:Ach_1_6}
&=\log\frac{{\rm{det}}(\bI_{l}+2\bar{P}\GcH\Gc)}{{\rm{det}}(\bI_{l}+\bar{P}\GcH\Gc)}\\
\label{eq:Ach_1_7}
&=\log\frac{2^{l}{\rm{det}}(\frac{1}{2}\bI_{l}+\bar{P}\GcH\Gc)}{{\rm{det}}(\bI_{l}+\bar{P}\GcH\Gc)}\\
\label{eq:Ach_1_8}
&\leq l.
\end{align}

Substituting (\ref{eq:Ach_1_2}) and (\ref{eq:Ach_1_8}) in (\ref{eq:AchSecRate}), we have 
\begin{align}
\label{eq:Ach_1_9}
R_s&\geq d\log P+o(\log P)-l\\
\label{eq:Ach_1_10}
&=(N+N_c-N_e)\log P+o(\log P)-N_c,
\end{align}
and hence, using (\ref{eq:sdof}), we conclude that the achievable s.d.o.f. is $D_s\geq N+N_c-N_e$.

\subsection{Case 2: $N_e\leq N, \frac{N_e}{2}<N_c\leq N$, and $N_e$ is even}{\label{AchScheme2}}
Unlike case $1$, the s.d.o.f. for this case does not increase by increasing $N_c$. For all $N_c$ in this case, the transmitter sends the same number of information streams, while the cooperative jammer utilizes a linear precoder which allows for discarding any unnecessary antennas. The s.d.o.f. here is integer valued, and we use Gaussian signaling for transmission and cooperative jamming. 

In particular, for $N_e$ is even, $N_c=\frac{N_e}{2}$, and $N_e\leq N$, the achievable s.d.o.f., using the scheme in Section \ref{AchScheme1}, is equal to $N-\frac{N_e}{2}$. However, from (\ref{eq:conv_3_1}), we observe that the s.d.o.f. is upper bounded by $N-\frac{N_e}{2}$ for all $\frac{N_e}{2}<N_c\leq {N}$. Thus, when $N_e\leq N$ and $N_e$ is even, the scheme for $N_c=\frac{N_e}{2}$ in Section \ref{AchScheme1} can be used to achieve the s.d.o.f. for all $\frac{N_e}{2}<N_c\leq N$, where the cooperative jammer uses the precoder  
\begin{align}
\label{eq:Ach_2_1}
\Pc=\begin{bmatrix}\bI_{l}\\\bz_{(N_c-l)\times l}\\ \end{bmatrix},
\end{align}
with $l=\frac{N_e}{2}$, to utilize only $\frac{N_e}{2}$ out of its $N_c$ antennas, and the transmitter utilizes
\begin{align}
\label{eq:Ach_2_2}
\Pt=\left[\Pta\;\Ptn\right],
\end{align}
$\Pta=\Gtpinv\Gc\Pc\in\mathbb{C}^{N\times l}$, $\Ptn\in\mathbb{C}^{N\times (N-N_e)}$ is defined as in (\ref{eq:Pt}), in order to send $d=N-\frac{N_e}{2}$ Gaussian information streams. Following the same analysis as in the previous case, the achievable s.d.o.f. is $N-\frac{N_e}{2}$ for all $\frac{N_e}{2}<N_c\leq N$, where $N_e$ is even and $N_e\leq N$. 

\subsection{Case 3: $N_e\leq N$, $\frac{N_e}{2}< N_c \leq N$, and $N_e$ is odd}{\label{AchScheme3}}
The s.d.o.f. for this case is equal to $N-\frac{N_e}{2}$, which is not an integer. As Gaussian signaling can not achieve fractional s.d.o.f. for the channel, we utilize structured signaling both for transmission and cooperative jamming for this case. In particular, we propose utilizing {\it{joint}} signal space alignment and the complex field equivalent of real interference alignment \cite{maddah2010degrees,kleinbock2002baker}. 

The decoding scheme at the receiver is as follows. The receiver projects its received signal over a direction that is orthogonal to all but one information and one cooperative jamming streams. Then, the receiver decodes these two streams from the projection using complex field analogy of real interference alignment. Finally, the receiver removes the decoded information and cooperative jamming streams from its received signal, leaving $N-1$ spatial dimensions for the other $N-\frac{N_e+1}{2}$ information and $\frac{N_e-1}{2}$ cooperative jamming streams. 

The transmitted signals are given by (\ref{eq:Xt1_Xc1}), with $d=N-\frac{N_e-1}{2}$, $l=\frac{N_e+1}{2}$, $\Pc, \Pt$ are defined as in (\ref{eq:Ach_2_1}) and (\ref{eq:Ach_2_2}), and $U_i=\Uire+j\Uiim$, $V_k=\Vkre+j\Vkim$, $i=2,3,\cdots,d$ and $k=2,3,\cdots,l$. The random variables $U_1$, $V_1$, $\{\Uire\}_{i=2}^{d}$, $\{\Uiim\}_{i=2}^{d}$, $\{\Vire\}_{i=2}^{l}$, and $\{\Viim\}_{i=2}^{l}$ are i.i.d. uniform over the set $\left\{a(-Q,Q)_{\mathbb{Z}}\right\}$. The values for $a$ and the integer $Q$ are chosen as 
\begin{align}
\label{eq:Q}
Q&=\left\lfloor P^{\frac{1-\epsilon}{2+\epsilon}}\right\rfloor =P^{\frac{1-\epsilon}{2+\epsilon}}-\nu\\
\label{eq:a}
a&=\gamma P^{\frac{3\epsilon}{2(2+\epsilon)}},
\end{align}
in order to satisfy the power constraints, where $\epsilon$ is an arbitrarily small positive number, and $\nu,\gamma$ are constants that do not depend on the power $P$. Justification for the choice of $a$ and $Q$ is provided in Appendix D. 

The received signal at the eavesdropper is  
\begin{align}
\label{eq:Ach_3_3}
\Ye&=\Gctilde(\Utonel+\Vc)+\Ze,
\end{align}
where $\Gctilde=\Gc\Pc$. We upper bound the second term in (\ref{eq:AchSecRate}), $I(\Xt;\Ye)$, as follows: 
\begin{align}
\label{eq:Ach_3_4}
I(\Xt;\Ye)&\leq I(\Xt;\Ye,\Ze)\\
\label{eq:Ach_3_5}
&=I(\Xt;\Ye|\Ze)\\
\label{eq:Ach_3_6}
&=H(\Ye|\Ze)-H(\Ye|\Ze,\Xt)\\
\label{eq:Ach_3_7}
&=H\left(\Gctilde(\Utonel+\Vc)\right)-H\left(\Gctilde\Vc\right)\\
\label{eq:Ach_3_8}
&=H(\Utonel+\Vc)-H(\Vc)\\
\label{eq:Ach_3_9}
&\nonumber =H\left(U_1+V_1,\Utwore+\Vtwore,\Utwoim+\Vtwoim,\cdots,\Ulre+\Vlre,\Ulim+\Vlim\right)\\
&\qquad\qquad\qquad-H\left(V_1,\Vtwore,\Vtwoim,\cdots,\Vlre,\Vlim\right)\\
\label{eq:Ach_3_10}
&\leq \log (4Q+1)^{2l-1}-\log(2Q+1)^{2l-1}\\
\label{eq:Ach_3_11}
&=(2l-1)\log\frac{4Q+1}{2Q+1}\\
\label{eq:Ach_3_12}
&\leq 2l-1,
\end{align}
where (\ref{eq:Ach_3_5}) follows since $\Xt$ and $\Ze$ are independent, and (\ref{eq:Ach_3_10}) follows since the entropy of a uniform random variable over the set $\left\{a(-2Q,2Q)_{\mathbb{Z}}\right\}$ upper bounds the entropy of each of $U_1+V_1,\Utwore+\Vtwore,\Utwoim+\Vtwoim,\cdots, \Ulim+\Vlim$. Equation (\ref{eq:Ach_3_8}) follows since the mappings $\Utonel+\Vc\mapsto\Gctilde(\Utonel+\Vc)$ and $\Vc\mapsto\Gctilde\Vc$ are bijective. The reason for this is that the entries of $\Gctilde$ are {\it{rationally independent}}, and that $(\Utonel+\Vc)$, $\Vc$ belong to $\mathbb{Z}^l[j]$.

\begin{Definition}\label{definition2}
A set of complex numbers $\{c_1,c_2,\cdots,c_L\}$ are rationally independent, i.e., linearly independent over $\mathbb{Q}$, if there is no set of rational numbers $\{r_i\}$, $r_i\neq 0$ for all $i=1,2,\cdots,L$, such that $\sum_{i=1}^L r_i c_i=0$.
\end{Definition}

Next, we derive a lower bound for $I(\Xt;\Yr)$. The received signal at the legitimate receiver is given by
\begin{align}
\label{eq:Ach_3_14}
\Yr&=\A\Ut+\Hcdash\Vc+\Zr,
\end{align}
where $\A=\Ht\Pt=\left[\ba_1\;\ba_2\;\cdots\;\ba_{d}\right]$ and $\Hcdash=\Hc\Pc=\left[\bh_{c,1}\;\bh_{c,2}\;\cdots\;\bold{h}_{c,l}\right]$. The receiver chooses $\bb\in\mathbb{C}^{N}$ such that $\bb\perp{\rm{span}}\left\{\ba_2,\cdots,\ba_d,\bh_{c,2},\cdots,\bold{h}_{c,l}\right\}$ and obtains 
\begin{align}
\Yrtilde=\bold{D}\Yr
\end{align}
where
\begin{align}
\label{eq:D}
\bold{D}=\begin{bmatrix}
\qquad\quad\bbH\\\bold{0}_{(N-1)\times 1}&\bold{I}_{N-1}
\end{bmatrix}.
\end{align}

Due to the fact that channel gains are continuous and randomly generated, $\ba_1$ and $\bh_{c,1}$ are linearly independent from ${\rm{span}}\left\{\ba_2,\cdots,\ba_d,\bh_{c,2},\cdots,\bold{h}_{c,l}\right\}$, and hence, $\bb$ is not orthogonal to $\ba_1$ and $\bh_{c,1}$ a.s. Thus, we have
\begin{align}
\label{eq:Ach_3_15}
\Yrtilde=\begin{bmatrix}\Yronetd\\\YrtwoNtd\\\end{bmatrix}
=\begin{bmatrix}\bbH\ba_1\;\;\bz_{1\times (d-1)}\\\Atilde\\\end{bmatrix}\begin{bmatrix}U_1\\\Uttwod\\\end{bmatrix}+
\begin{bmatrix}\bbH\bh_{c,1}\;\;\bz_{1\times (l-1)}\\\Hctilde\\\end{bmatrix}\begin{bmatrix}V_1\\\Vctwol\\\end{bmatrix}+
\begin{bmatrix}\bbH\Zr\\\ZrtwoN\\\end{bmatrix},
\end{align}
where $\Atilde=\left[\tba_1\;\tba_2\;\cdots\;\tba_d\right]\in\mathbb{C}^{(N-1)\times d}$, $\tba_i={\ba_{i}}_2^N$ for all  $i=1,2,\cdots,d$. Similarly, $\Hctilde=[\tbh_{c,1}\;\tbh_{c,2}\;\cdots\;\tbh_{c,{l}}]\in\mathbb{C}^{(N-1)\times l}$, where $\tbh_{c,2}={\bh_{c,i}}_2^N$ for all $i=1,2,\cdots,l$. 

Next, the receiver uses $\Yronetd$ to decode the information stream $U_1$ and the cooperative jamming stream $V_1$ as follows. Let $Z'=\bbH\Zr\sim\mathcal{CN}(0,||\bb||^2)$, $f_1=\bbH\ba_1$, and $f_2=\bbH\bh_{c,1}$. Thus, $\Yronetd$ is given by
\begin{align}
\label{eq:Ach_3_16}
\Yronetd=f_1U_1+f_2V_1+Z'.
\end{align}
Once again, with randomly generated channel gains, $f_1=\bbH\ba_1$ and $f_2=\bbH\bh_{c,1}$ are   rationally independent a.s. Thus, the mapping $(U_1,V_1)\mapsto f_1U_1+f_2V_1$ is invertible \cite{motahari2009real2}. The receiver employs a hard decision decoder which maps $\Yronetd\in\tilde{\mathcal{Y}}_{r_1}$ to the nearest point in the constellation $\mathcal{R}_1=f_1\mathcal{U}_1+f_2\mathcal{V}_1$, where $\mathcal{U}_1,\mathcal{V}_1=\left\{a(-Q,Q)_{\mathbb{Z}}\right\}$. Then, the receiver passes the output of the hard decision decoder through the bijective mapping $f_1U_1+f_2V_1\mapsto(U_1,V_1)$ in order to decode both $U_1$ and $V_1$. 

The receiver can now use
\begin{align}
\label{eq:Ach_3_17}
\Yrbar&=\YrtwoNtd-\tba_1 U_1-\tbh_{c,1} V_1\\
\label{eq:Ach_3_18}
&=\begin{bmatrix}\tba_2&\cdots&\tba_{d}\\\end{bmatrix}\Uttwod+\begin{bmatrix}\tbh_{c,2}&\cdots&\tbh_{c,l}\\\end{bmatrix}\Vctwol+\ZrtwoN\\
\label{eq:Ach_3_19}
&=\bold{B} \begin{bmatrix}\Uttwod\\\Vctwol\\\end{bmatrix}+\ZrtwoN,
\end{align}
to decode $U_2,\cdots,U_d$, where, 
\begin{align}
\label{eq:Ach_3_20}
\bold{B}=\left[\tba_2\;\cdots\;\tba_{d}\;\;\tbh_{c,2}\;\cdots\;\tbh_{c,l}\right]\in\mathbb{C}^{(N-1)\times (N-1)},
\end{align}
is full rank a.s.  To show that $\bold{B}$ is full rank a.s., let $\Htbar$ and $\Hcbar$ be generated by removing the first row from $\Ht$ and $\Hc$, and let $\Ptbar$ and $\Pcbar$ be generated by removing the first column from $\Pt$ and $\Pc$, respectively. $\bold{B}$ can be rewritten as 
\begin{align}
\label{eq:Ach_3_20_1}
\bold{B}=\begin{bmatrix}\Htbar&\Hcbar\\\end{bmatrix}\begin{bmatrix}\Ptbar&\bz_{N\times (l-1)}\\\bz_{N_c\times (d-1)}&\Pcbar\\\end{bmatrix}.
\end{align}
Note that $\left[\Htbar\;\Hcbar\right]$ has all of its entries independently and randomly drawn from a continuous distribution, and the second matrix in the RHS of (\ref{eq:Ach_3_20_1}) is full column rank. Using Lemma \ref{lemma1}, the matrix $\bold{B}$ is full rank a.s.

Hence, by zero forcing, the receiver obtains
\begin{align}
\label{eq:Ach_3_21}
\Yrhat&=\bold{B}^{-1}\Yrbar=\begin{bmatrix}\Uttwod\\\Vctwol\\\end{bmatrix}+\bar{\bold{Z}}_r,
\end{align}
where $\bar{\bold{Z}}_r=\bold{B}^{-1}\ZrtwoN \sim\mathcal{CN}\left(\bz,\bold{B}^{-1}\bold{B}^{-H}\right)$. Thus, at high SNR, the receiver can decode the other information streams, $U_2,\cdots,U_d$, from $\Yrhat$.

The mutual information between the transmitter and receiver is lower bounded as follows:
\begin{align}
\nonumber \\
\label{eq:Ach_3_22}
I(\Xt;\Yr)&\geq I(\Ut;\Yrtilde)\\
\label{eq:Ach_3_23}
&=I(U_1,\Uttwod;\Yronetd,\YrtwoNtd)\\
\label{eq:Ach_3_24}
&=I(U_1,\Uttwod;\Yronetd)+I(U_1,\Uttwod;\YrtwoNtd|\Yronetd)\\
\label{eq:Ach_3_25}
&=I(U_1;\Yronetd)+I(\Uttwod;\Yronetd|U_1)+I(U_1;\YrtwoNtd|\Yronetd)+I(\Uttwod;\YrtwoNtd|U_1,\Yronetd)\\
\label{eq:Ach_3_26}
&\geq I(U_1;\Yronetd)+I(\Uttwod;\YrtwoNtd|U_1,\Yronetd),
\end{align}
where (\ref{eq:Ach_3_22}) follows since $\Ut-\Xt-\Yr-\Yrtilde$ forms a Markov chain. We next lower bound each term in the RHS of (\ref{eq:Ach_3_26}).

We lower bound the first term, $I(U_1;\Yronetd)$ as follows, see also \cite{motahari2009real2,maddah2010degrees}. Let $P_{e_1}$ denote the probability of error in decoding $U_1$ at the receiver, i.e., $P_{e_1}=\Pro\left\{\hat{U}_1\neq U_1\right\}$, where $\hat{U}_i$, $i=1,2,\cdots,d$, is the estimate of $U_i$ at the legitimate receiver. Thus, using Fano's inequality, we have 
\begin{align}
\label{eq:Ach_3_27}
I(U_1;\Yronetd)&=H(U_1)-H(U_1|\Yronetd)\\
\label{eq:Ach_3_28}
& \geq H(U_1)-1-P_{e_1}\log |\mathcal{U}_1|\\
\label{eq:Ach_3_29}
&=\left(1-P_{e_1}\right)\log(2Q+1)-1.
\end{align}
From (\ref{eq:Ach_3_16}), since the mapping $(U_1,V_1)\mapsto f_1 U_1+f_2V_1$ is invertible, the only source of error in decoding $U_1$ from $\Yronetd$ is the additive Gaussian noise $Z'$. Note that, since $Z'\sim\mathcal{CN}\mathcal(0,||\bb||^2)$, $\Real\{Z'\}$ and $\Ima\{Z'\}$ are i.i.d. with $\mathcal{N}\left(0,\frac{||\bb||^2}{2}\right)$ distribution, and $|Z'|\sim {\rm{Rayleigh}}\left(\frac{||\bb||}{\sqrt{2}}\right)$. Thus, we have 
\begin{align}
\label{eq:Ach_3_30}
P_{e_1}&=\Pro\left\{\hat{U}_1\neq U_1\right\}\\
\label{eq:Ach_3_31}
&\leq \Pro\left\{(\hat{U}_1,\hat{V}_1)\neq (U_1,V_1)\right\}\\
\label{eq:Ach_3_32}
&\leq \Pro\left\{|Z'|\geq \frac{\dmin}{2}\right\}\\
\label{eq:Ach_3_33}
&=\exp\left(\frac{-\dmin^2}{4||\bb||^2}\right),
\end{align}
where $\dmin$ is the minimum distance between the points in the constellation $\mathcal{R}_1=f_1\mathcal{U}_1+f_2\mathcal{V}_1$. 

In order to upper bound $P_{e_1}$, we lower bound $\dmin$. To do so, similar to \cite{maddah2010degrees}, we extend real interference alignment \cite{motahari2009real2} to complex channels. In particular, we utilize the following results from number theory:

\begin{Definition}\label{definition1}\cite{kleinbock2002baker}
The Diophantine exponent $\omega(\bold{z})$ of $\bold{z}\in\mathbb{C}^n$ is defined as 
\begin{align}
\label{eq:Ach_3_34}
\omega(\bold{z})=\sup\left\{v:|p+\bold{z}.\bold{q}|\leq (||\bold{q}||_{\infty})^{-v} \text{ for infinetly many } \bold{q}\in\mathbb{Z}^n, p\in\mathbb{Z}\right\},
\end{align}
where $\bold{q}=[q_1\;q_2\;\cdots\;q_n]^T$ and $||\bold{q}||_\infty=\underset{i}\max|q_i|$.
\end{Definition}

\begin{lemma}\label{lemma2}\cite{kleinbock2002baker}
For almost all $\bold{z}\in\mathbb{C}^n$, the Diophantine exponent $\omega(\bold{z})$ is equal to $\frac{n-1}{2}$.
\end{lemma}
Lemma \ref{lemma2} implies the following:
\begin{corollary}\label{Corollary1}
For almost all $\bold{z}\in\mathbb{C}^n$ and for all $\epsilon>0$, 
\begin{align}
\label{eq:Ach_3_35}
|p+\bold{z}.\bold{q}|>(\underset{i}\max|q_i|)^{-\frac{(n-1+\epsilon)}{2}},
\end{align}
holds for all $\bold{q}\in\mathbb{Z}^n$ and $p\in\mathbb{Z}$ except for finitely many of them.  
\end{corollary}
Since the number of integers that violate the inequality in (\ref{eq:Ach_3_35}) is finite, there exists a constant $\kappa$ such that, for almost all $\bold{z}\in\mathbb{C}^n$ and all $\epsilon>0$, the inequality
\begin{align}
\label{eq:Ach_3_35_1}
|p+\bold{z}.\bold{q}|>\kappa(\underset{i}\max|q_i|)^{-\frac{(n-1+\epsilon)}{2}},
\end{align}
holds for all $\bold{q}\in\mathbb{Z}^n$ and $p\in\mathbb{Z}$.

Thus, for almost all channel gains, the minimum distance $\dmin$ is lower bounded as follows: 
\begin{align}
\label{eq:Ach_3_36}
\dmin&=\inf_{Y'_{r_1}, Y''_{r_1}\in\mathcal{R}_1} |Y'_{r_1}-Y''_{r_1}|\\
\label{eq:Ach_3_37}
&=\inf_{U_1, V_1\in\left\{a(-2Q,2Q)_{\mathbb{Z}}\right\}} |f_1 U_1+f_2 V_1|\\
\label{eq:Ach_3_38}
&=\inf_{U_1, V_1\in(-2Q,2Q)_{\mathbb{Z}}} a |f_1| \left|U_1+\frac{f_2}{f_1} V_1\right|\\
\label{eq:Ach_3_39}
&\geq \kappa\frac{a |f_1|}{(2Q)^{\frac{\epsilon}{2}}}\\
\label{eq:Ach_3_40}
&\geq \kappa\gamma |f_1| 2^{-\frac{\epsilon}{2}} P^{\frac{\epsilon}{2}},
\end{align}
where (\ref{eq:Ach_3_39}) follows from (\ref{eq:Ach_3_35_1}), and (\ref{eq:Ach_3_40}) follows by substituting (\ref{eq:Q}) and (\ref{eq:a}) in (\ref{eq:Ach_3_39}). Substituting (\ref{eq:Ach_3_40}) in (\ref{eq:Ach_3_33}) gives the following bound on $P_{e_1}$,
\begin{align}
\label{eq:Ach_3_41}
P_{e_1}\leq \exp(-\mu P^{\epsilon}),
\end{align} 
where $\mu=\frac{\kappa^2\gamma^2 |f_1|^2 2^{-\epsilon}}{4||\bb||^2}$ is a constant which does not depend on the power $P$. Thus, using (\ref{eq:Ach_3_29}) and (\ref{eq:Ach_3_41}), we have 
\begin{align}
\label{eq:Ach_3_42}
I(U_1;\Yronetd)\geq \left(1-\exp(-\mu P^{\epsilon})\right)\log(2Q+1)-1.
\end{align} 

Next, we lower bound the second term in the RHS of (\ref{eq:Ach_3_26}), $I(\Uttwod;\YrtwoNtd|U_1,\Yronetd)$. Let $\widetilde{\bold{B}}=\begin{bmatrix}\bz_{(N-1) \times 1}&\bI_{N-1}\\\end{bmatrix}-\frac{1}{f_2}\tbh_{c,1}\bbH$, and 
\begin{align}
\label{eq:Ach_3_43}
\Yrbardash &=\bold{B}\begin{bmatrix}\Uttwod\\\Vctwol\\\end{bmatrix}+\widetilde{\bold{B}}\Zr\\
\label{eq:Ach_3_44}
\Yrhatdash &=\bold{B}^{-1}\Yrbardash= \begin{bmatrix}\Uttwod\\\Vctwol\\\end{bmatrix}+\bold{B}^{-1}\widetilde{\bold{B}}\Zr,
\end{align}
where $\bold{B}$ is defined as in (\ref{eq:Ach_3_20}). Thus, we have
\begin{align}
\label{eq:Ach_3_45}
I\left(\Uttwod;\YrtwoNtd|U_1,\Yronetd\right)&=I\left(\Uttwod;\Atilde\Ut+\Hctilde\Vc+\ZrtwoN \big|U_1,f_2 V_1+Z'\right)\\
\label{eq:Ach_3_46}
&=I\left(\Uttwod;\bold{B}\begin{bmatrix}\Uttwod\\\Vctwol\\\end{bmatrix}+\ZrtwoN-\frac{1}{f_2}\tbh_{c,1}\bbH\Zr\bigg|f_2 V_1+Z'\right)\\
\label{eq:Ach_3_47}
&=I(\Uttwod;\Yrbardash|f_2 V_1+Z')\\
\label{eq:Ach_3_48}
&\geq I(\Uttwod;\Yrbardash)\\
\label{eq:Ach_3_49}
&\geq I(\Uttwod;\Yrhatdash)\\
\label{eq:Ach_3_50}
&=H(\Uttwod)-H(\Uttwod|\Yrhatdash)\\
\label{eq:Ach_3_51}
&\geq H(\Uttwod)-P_{e_2}^d\log(2Q+1)^{2(d-1)}-1\\
\label{eq:Ach_3_52}
&= 2(d-1)\left(1-P_{e_2}^d\right)\log(2Q+1)-1,
\end{align}
where $P_{e_2}^d=\Pro\{(\hat{U}_2,\hat{U}_3,\cdots,\hat{U}_d)\neq (U_2,U_3,\cdots,U_d)\}$, (\ref{eq:Ach_3_45}) follows from (\ref{eq:Ach_3_15}), (\ref{eq:Ach_3_48}) follows since $\Uttwod$ and $f_2 V_1+Z'$ are independent, (\ref{eq:Ach_3_49}) follows since $\Uttwod-\Yrbardash-\Yrhatdash$ forms a Markov chain, and (\ref{eq:Ach_3_51}) follows from Fano's inequality. 

Let $\widehat{\bold{Z}}_r=\bold{\Theta}\Zr=[\hat{Z}_{r_2}\;\cdots\;\hat{Z}_{r_N}]^T$, where $\bold{\Theta}=\bold{B}^{-1}\widetilde{\bold{B}}$. Thus, $\widehat{\bold{Z}}_r\sim\mathcal{CN}(\bz,\bold{\Theta}\bold{\Theta}^H)$ and $|\hat{Z}_{r_i}|\sim {\rm{Rayleigh}}(\sigma_i)$, where $\sigma_i^2=\bold{\Theta}\bold{\Theta}^H(i,i)$, $i=2,3,\cdots,N$. Using the union bound, we have 
\begin{align}
\label{eq:Ach_3_53}
P_{e_2}^d&=\Pro\left\{(\hat{U}_2,\hat{U}_3,\cdots,\hat{U}_d)\neq (U_2,U_3,\cdots,U_d)\right\}\\
\label{eq:Ach_3_54}
&\leq \sum_{i=2}^{d}\Pro\left\{\hat{U}_i\neq U_i\right\}\\
\label{eq:Ach_3_55}
&\leq\sum_{i=2}^d \Pro\left\{|\hat{Z}_{r_i}|\geq \frac{a}{2}\right\}\\
\label{eq:Ach_3_56}
&=\sum_{i=2}^d \exp\left(-\frac{a^2}{8\sigma_i^2}\right)\\
\label{eq:Ach_3_57}
&\leq (d-1)\exp\left(-\frac{\gamma^2}{8\sigma_{\rm{max}}^2}P^{\frac{3\epsilon}{2+\epsilon}}\right)\\
\label{eq:Ach_3_58}
&=\left(d-1\right)\exp(-\mu' P^{\epsilon'}),
\end{align}
where $\sigma_{\rm{max}}=\underset{i}\max\;\sigma_i$, $\mu'=\frac{\gamma^2}{8\sigma_{\rm{max}}^2}$, $\epsilon'=\frac{3\epsilon}{2+\epsilon}$, and (\ref{eq:Ach_3_57}) follows by substituting (\ref{eq:a}) in (\ref{eq:Ach_3_56}).

Substituting (\ref{eq:Ach_3_58}) in (\ref{eq:Ach_3_52}) yields
\begin{align}
\label{eq:Ach_3_59}
I\left(\Uttwod;\YrtwoNtd|U_1,\Yronetd\right) &\geq \left(2d-2-2(d-1)^2\exp(-\mu' P^{\epsilon'})\right)\log(2Q+1)-1.
\end{align}
Using (\ref{eq:Q}), (\ref{eq:Ach_3_26}), (\ref{eq:Ach_3_42}), and (\ref{eq:Ach_3_59}), we have  
\begin{align}
\label{eq:Ach_3_60}
I(\Xt&;\Yr)\geq \left[2d-1-\exp(-\mu P^{\epsilon})-2(d-1)^2\exp(-\mu' P^{\epsilon'})\right]\log(2P^{\frac{1-\epsilon}{2+\epsilon}}-2\nu+1)-2\\
\label{eq:Ach_3_61}
&=\frac{1-\epsilon}{2+\epsilon}\left[2d-1-\exp(-\mu P^{\epsilon})-2(d-1)^2\exp(-\mu' P^{\epsilon'})\right]\log P+o(\log P).
\end{align}

Using the upper bound in (\ref{eq:Ach_3_12}) and the lower bound in (\ref{eq:Ach_3_61}), we get 
\begin{align}
\label{eq:Ach_3_62}
&R_s \geq \frac{1-\epsilon}{2+\epsilon}\left[2d-1-\exp(-\mu P^{\epsilon})-2(d-1)^2\exp(-\mu' P^{\epsilon'})\right]\log P+o(\log P)-(2l-1)\\
&=\frac{1-\epsilon}{2+\epsilon}\left[2N-N_e-\exp(-\mu P^{\epsilon})-\frac{1}{2}(2N-N_e-1)^2\exp(-\mu' P^{\epsilon'})\right]\log P+o(\log P)-N_e.
\end{align}
Thus, it follows that the s.d.o.f. is lower bounded as 
\begin{align}
\label{eq:Ach_3_63}
D_s \geq \frac{(1-\epsilon)(2N-N_e)}{2+\epsilon}. 
\end{align}
Since $\epsilon>0$ can be chosen arbitrarily small, we can achieve s.d.o.f. of $N-\frac{N_e}{2}$.

\subsection{Case 4: $N_e\leq N$, $N<N_c\leq N+N_e$, and $N+N_c-N_e$ is even}{\label{AchScheme4}}
Since $N_c>N$ for this case, the cooperative jammer, unlike the previous three cases, chooses its precoder such that $N_c-N$ of its jamming streams are sent invisible to the receiver, in order to allow for more space for the information streams at the receiver. The s.d.o.f. for this case is integer valued, which we can achieve using Gaussian information and cooperative jamming streams. 

The transmitted signals are given by (\ref{eq:Xt1_Xc1}), with $d=\frac{N+N_c-N_e}{2}$, $l=\frac{N_c+N_e-N}{2}$, $\Ut\sim\mathcal{CN}\left(\bz,\bar{P}\bI_d\right)$, $\Vc\sim\mathcal{CN}\left(\bz,\bar{P}\bI_l\right)$, 
\begin{align}
\label{eq:Ach_4_0}
\Pc=[\PcI\;\Pcn],
\end{align} 
where $\PcI$ is given by
\begin{align}
\label{eq:Ach_4_1}
\PcI=\begin{bmatrix}\bI_{g}\\\bz_{(N_c-g)\times g}\\\end{bmatrix},
\end{align}
$g=\frac{N_e+N-N_c}{2}$, and $\Pcn\in\mathbb{C}^{N_c\times (N_c-N)}$ is a matrix whose columns span $\mathcal{N}(\Hc)$, $\Pt$ is defined as in Section \ref{AchScheme2}, and $\bar{P}=\frac{1}{\alpha'}P$, where $\alpha'=\max\left\{\sum_{i=1}^d||\bp_{t,i}||^2, g+\sum_{i=g+1}^{l}||\bp_{c,i}||^2\right\}$. At high SNR, the receiver can decode the $d$ information and the $g$ cooperative jamming streams, where $d+g=N$. 

The received signals at the legitimate receiver and the eavesdropper are given by 
\begin{align}
\label{eq:Yr_4_1}
\Yr&=\Ht\Pt\Ut+\begin{bmatrix}\Hc\PcI&\bz_{N\times (N_c-N)}\\\end{bmatrix}\begin{bmatrix}\Vconeg\\\Vcgponel\\\end{bmatrix}+\Zr\\
\label{eq:Yr_4_2}
&=\begin{bmatrix}\Ht\Pt&\Hc\PcI\\\end{bmatrix}\begin{bmatrix}\Ut\\\Vconeg\\\end{bmatrix}+\Zr\\
\label{eq:Ye_4_1}
\Ye&=\Gctilde(\Utonel+\Vc)+\Ze,
\end{align}
where $\Gctilde=\Gc\Pc$. 

The matrix $\left[\Ht\Pt\;\;\Hc\PcI\right]\in\mathbb{C}^{N\times N}$ in (\ref{eq:Yr_4_2}) can be rewritten as
\begin{align}
\label{eq:Ach_4_2}
\begin{bmatrix}\Ht\Pt&\Hc\PcI\\\end{bmatrix}=\begin{bmatrix}\Ht&\Hc\\\end{bmatrix}\begin{bmatrix}\Pt&\bz_{N\times g}\\\bz_{N_c\times d}&\PcI\\\end{bmatrix}.
\end{align}
By applying Lemma \ref{lemma1} on (\ref{eq:Ach_4_2}), the matrix $\left[\Ht\Pt\;\;\Hc\PcI\right]$ is full rank a.s. Thus, 
\begin{align}
I(\Xt;\Yr)\geq d\log P+o(\log P).
\end{align}

Using similar steps as from (\ref{eq:Ach_1_3}) to (\ref{eq:Ach_1_8}), we can show that
\begin{align}
\label{eq:Ach_4_4}
I(\Xt;\Ye)&= \log\frac{\det(\bI_l+2\bar{P}\GctildeH\Gctilde)}{\det(\bI_l+\bar{P}\GctildeH\Gctilde)}\leq l.
\end{align}
Thus, the achievable secrecy rate in (\ref{eq:AchSecRate}) is lower bounded as 
\begin{align}
\label{eq:Ach_4_4_1}
R_s&\geq d\log P+o(\log P)-l\\
\label{eq:Ach_4_5}
&=\frac{N+N_c-N_e}{2}\log P+o(\log P)-\frac{N_c+N_e-N}{2},
\end{align} 
and, using (\ref{eq:sdof}), the s.d.o.f. is lower bounded as  
\begin{align}
\label{eq:Ach_4_4}
D_s\geq \frac{N+N_c-N_e}{2}. 
\end{align}

\subsection{Case 5: $N_e\leq N$, $N<N_c\leq N+N_e$, and $N+N_c-N_e$ is odd}{\label{AchScheme5}}
As in case $3$, the s.d.o.f. for this case is not an integer, and as in case $4$, we have $N_c>N$, which allows the cooperative jammer to send some signals invisible to the receiver. Consequently, the achievable scheme for this case combines the techniques used in Sections \ref{AchScheme3} and \ref{AchScheme4}.  

The transmitted signals are given by (\ref{eq:Xt1_Xc1}) with $d=\frac{N+N_c-N_e+1}{2}$, $l=\frac{N_c+N_e-N+1}{2}$, $\Pt$ and $\Pc$ are defined as in Section \ref{AchScheme4} with $g=\frac{N_e+N-N_c+1}{2}$, and $\Ut$, $\Vc$ are defined as in Section \ref{AchScheme3}. Similar to the proof in Appendix D, the values of $Q$ and $a$ are chosen as in (\ref{eq:Q}) and (\ref{eq:a}), with     
\begin{align}
\gamma=\frac{1}{\sqrt{\max\left\{\||\bp_{t,1}||^2+2\sum_{i=2}^d||\bp_{t,i}||^2, 2g-1+2\sum_{i=g+1}^l||\bp_{c,i}||^2\right\}}},
\end{align}
and $\nu$ are constants that do not depend on the power $P$.

The legitimate receiver uses the projection and cancellation technique described in Section \ref{AchScheme3} in order to decode the information streams. The received signal at the eavesdropper is the same as in (\ref{eq:Ye_4_1}), with $l=\frac{N_c+N_e-N+1}{2}$. Similar to the derivation from (\ref{eq:Ach_3_4}) to (\ref{eq:Ach_3_12}), we have 
\begin{align}
\label{eq;Ach_5_1}
I(\Xt;\Ye)\leq 2l-1.
\end{align}

Let $\A=\Ht\Pt=\left[\ba_1\cdots\ba_d\right]$, and $\Hcdash=\Hc\PcI=[\bh_{c,1}\cdots\bh_{c,g}]$. The received signal at the legitimate receiver is
\begin{align}
\label{eq:Yr_5_1}
\Yr=\begin{bmatrix}\A&\Hcdash\\\end{bmatrix}\begin{bmatrix}\Ut\\\Vconeg\\\end{bmatrix}+\Zr.
\end{align}
The receiver chooses $\bb\perp{\rm{span}}\left\{\ba_2,\cdots,\ba_d,\bh_{c_2},\cdots,\bh_{c_g}\right\}$ and multiplies its received signal by the matrix $\bold{D}$ given in (\ref{eq:D}) to obtain $\Yrtilde=\left[\Yronetd\;(\YrtwoNtd)^T\right]^T$, where
\begin{align}
\label{eq;Ach_5_2}
\Yronetd&=f_1 U_1+f_2 V_1+Z',\\
\label{eq;Ach_5_3}
\YrtwoNtd&=\Atilde\Ut+\Hctilde\Vconeg+\ZrtwoN,
\end{align}
$f_1$, $f_2$, $Z'$, $\Atilde$, and $\Hctilde$, are defined as in Section \ref{AchScheme3}. In order to decode $U_1$ and $V_1$, the receiver passes $\Yronetd$ through a hard decision decoder, $\Yronetd \mapsto f_1 U_1+f_2 V_1$, and passes the output of the hard decision decoder through the bijective map $f_1 U_1+f_2 V_1\mapsto (U_1,V_1)$, where $f_1$ and $f_2$ are rationally independent. 

Using similar steps to the derivation from (\ref{eq:Ach_3_22}) to (\ref{eq:Ach_3_61}) in Section \ref{AchScheme3}, we obtain  
\begin{align}
\label{eq;Ach_5_6}
I(\Xt;\Yr)\geq \frac{1-\epsilon}{2+\epsilon}\left[2d-1-\exp\left(-\mu P^{\epsilon}\right)-2(d-1)^2\exp(-\mu' P^{\epsilon'})\right]\log P+o(\log P),
\end{align} 
where $\epsilon>0$ is arbitrarily small, $\epsilon'=\frac{3\epsilon}{2+\epsilon}$, and $\mu,\mu'$ are constants which do not depend on $P$.

Thus, the achievable secrecy rate in (\ref{eq:AchSecRate}) is lower bounded as 
\begin{align}
\label{eq;Ach_5_7_1}
&R_s\geq \frac{1-\epsilon}{2+\epsilon}\left[2d-1-\exp(-\mu P^{\epsilon})-(d-1)^2\exp(-\mu' P^{\epsilon'})\right]\log P+o(\log P)-(2l-1)\\
\label{eq;Ach_5_7}
\nonumber &= \frac{1-\epsilon}{2+\epsilon}\left[N+N_c-N_e-\exp(-\mu P^{\epsilon})-\frac{1}{2}(N+N_c-N_e-1)^2\exp(-\mu' P^{\epsilon'})\right]\log P\\
&\qquad\qquad\qquad +o(\log P)-(N_c+N_e-N),
\end{align}
and hence the s.d.o.f is lower bounded as 
\begin{align}
\label{eq:Ach_5_8}
D_s \geq \frac{(1-\epsilon)(N+N_c-N_e)}{2+\epsilon}.
\end{align}
Since $\epsilon>0$ can be chosen arbitrarily small, $D_s=\frac{N+N_c-N_e}{2}$ is achievable for this case, which completes the achievability of (\ref{eq:conv_3_1}). Next, we show the achievability of (\ref{eq:conv_3_2}), where $N_e>N$, i.e., the eavesdropper has more antennas than the legitimate receiver.
 
\subsection{Case 6: $N_e>N$ and $N_e-N<N_c\leq N_e-\frac{N}{2}$}{\label{AchScheme6}}
Unlike the previous five cases, since $N_e>N$, no information streams can be sent invisible to the eavesdropper. In fact, the precoder at the transmitter is not adequate for achieving the alignment of the information and cooperative jamming streams at the eavesdropper. We need to design both precoders at the transmitter and the cooperative jammer to take part in achieving the alignment condition. The s.d.o.f. here is integer valued, and hence we can utilize Gaussian streams. 

The transmitted signals are given by (\ref{eq:Xt1_Xc1}), with $d=l=N+N_c-N_e$, and $\Ut,\Vc\sim\mathcal{CN}\left(\bz,\bar{P}\bI_{d}\right)$. The matrices $\Pt$ and $\Pc$ are chosen as follows. Let $\bold{G}=\left[\Gt\;\;-\Gc\right]\in\mathbb{C}^{N_e\times (N+N_c)}$, and let $\bold{Q}\in\mathbb{C}^{(N+N_c)\times d}$ be a matrix whose columns are randomly\footnote{Out of all possible sets of $d=N+N_c-N_e$ linearly independent vectors which span $\mathcal{N}(\bold{G})$, the columns of $\bold{Q}$ are the elements of one randomly chosen set.} chosen to span $\mathcal{N}(\bold{G})$. Write the matrix $\bold{Q}$ as
$\bold{Q}=\left[\bold{Q}_1^T\;\;\bold{Q}_2^T\right]^T$, where $\bold{Q}_1\in\mathbb{C}^{N\times d}$ and $\bold{Q}_2\in\mathbb{C}^{N_c\times d}$. Set $\Pt=\bold{Q}_1$ and $\Pc=\bold{Q}_2$. $\bar{P}=\frac{1}{\alpha'' }P$, where $\alpha''=\max\left\{\sum_{i=1}^d ||\bp_{t,i}||^2,\sum_{i=1}^d||\bp_{c,i}||^2\right\}$.

The choice of $\Pt$ and $\Pc$ results in $\Gt\Pt=\Gc\Pc$. Thus, the eavesdropper receives
\begin{align}
\label{eq:Ach_6_1}
\Ye=\Gc\Pc(\Ut+\Vc)+\Ze.
\end{align}
Similar to going from (\ref{eq:Ach_1_3}) to (\ref{eq:Ach_1_8}), it follows that we have 
\begin{align}
\label{eq:Ach_6_2}
I(\Xt;\Ye)\leq N+N_c-N_e.
\end{align}

The received signal at the receiver in turn is given by 
\begin{align}
\label{eq:Ach_6_3}
\Yr=\begin{bmatrix}\Ht\Pt&\Hc\Pc\\\end{bmatrix}\begin{bmatrix}\Ut\\\Vc\\\end{bmatrix}+\Zr.
\end{align}
Note that, without conditioning on $\Gt$ and $\Gc$, the matrix $\bold{Q}$ has all of its entries independently and randomly drawn according to a continuous distribution. Thus, each of $\Pt$ and $\Pc$ is full column rank a.s. Thus, by using Lemma \ref{lemma1}, we can show that the matrix $\left[\Ht\Pt\;\;\Hc\Pc\right]$ is full column rank a.s. Using (\ref{eq:Ach_6_3}), we have
\begin{align}
\label{eq:Ach_6_4}
I(\Xt;\Yr)\geq (N+N_c-N_e)\log P+o(\log P).
\end{align} 
Hence, using (\ref{eq:Ach_6_2}), (\ref{eq:Ach_6_4}), (\ref{eq:AchSecRate}), and (\ref{eq:sdof}), the s.d.o.f. is lower bounded as $D_s\geq N+N_c-N_e$.

\subsection{Case 7: $N_e>N$, $N_e-\frac{N}{2}<N_c\leq N_e$, and $N$ is even}{\label{AchScheme7}}
The s.d.o.f. for this case does not increase by increasing $N_c$. The scheme in Section \ref{AchScheme6} for $N_c=N_e-\frac{N}{2}$, i.e., $d=\frac{N}{2}$, can be used to achieve the s.d.o.f. for all $N_e-\frac{N}{2}<N_c\leq N_e$, when $N_e>N$ and $N$ is even. However, since $\dim(\mathcal{N}(\bold{G}))=N+N_c-N_e>\frac{N}{2}$, the $d=\frac{N}{2}$ columns of the matrix $\bold{Q}$ are randomly chosen as linearly independent vectors from $\mathcal{N}(\bold{G})$. Following the same analysis as in Section \ref{AchScheme6}, we can show that the s.d.o.f. is lower bounded as $D_s\geq \frac{N}{2}$. 

\subsection{Case 8: $N_e>N$, $N_e-\frac{N}{2}<N_c\leq N_e$, and $N$ is odd}{\label{AchScheme8}}
The difference here from Section \ref{AchScheme7} is that s.d.o.f. is not an integer, and hence, structured signaling for transmission and cooperative jamming is needed, and the difference from \ref{AchScheme3} is that $N_e>N$, and hence both the precoders at the transmitter and cooperative jammer have to participate in achieving the alignment condition at the eavesdropper.

The transmitted signals are given by (\ref{eq:Xt1_Xc1}), with $d=l=\frac{N+1}{2}$, $\Ut$ and $\Vc$ are defined as in Section \ref{AchScheme3}, and the values for $Q$ and $a$ are chosen as in (\ref{eq:Q}) and (\ref{eq:a}), with 
\begin{align}
\label{eq:Ach_8_1}
\gamma=\frac{1}{\sqrt{\max\left\{\||\bp_{t,1}||^2+2\sum_{i=2}^d||\bp_{t,i}||^2, ||\bp_{c,1}||^2+2\sum_{i=2}^d||\bp_{c,i}||^2\right\}}},
\end{align}
and $\nu$ are constants which do not depend $P$. $\Pt,\Pc$ are chosen as in Section \ref{AchScheme7}, with $d=\frac{N+1}{2}$. The eavesdropper's received signal is the same as in (\ref{eq:Ach_6_1}). Similar to (\ref{eq:Ach_3_4})-(\ref{eq:Ach_3_12}), we have 
\begin{align}
\label{eq:Ach_8_2}
I(\Xt;\Ye)\leq N.
\end{align}

The receiver employs the decoding scheme in Sections \ref{AchScheme3} and \ref{AchScheme5}. Following similar steps as in Sections \ref{AchScheme3} and \ref{AchScheme5}, we have
\begin{align}
\label{eq:Ach_8_3}
I(\Xt;\Yr)\geq \frac{(1-\epsilon)N}{2+\epsilon}\log P+o(\log P).
\end{align}
Using (\ref{eq:Ach_8_2}), (\ref{eq:Ach_8_3}), (\ref{eq:AchSecRate}), and (\ref{eq:sdof}), the s.d.o.f. is lower bounded as $D_s\geq \frac{(1-\epsilon)N}{2+\epsilon}$, and since $\epsilon>0$ is arbitrarily small, the s.d.o.f. of $\frac{N}{2}$ is achievable for this case. 
                                                                                                                                                                                                                                                                                                                                                                                                                                                                                                                                                                                                                                                                                                                                                                                                                                                                                                                                                                                                                                                                                                                                                                                                                                                                                                                                                                                                                                                                                                                                                                                                                                                                                                                                                                                                                                                                                                                                                                                                                                                                                                                                                                                                                                                                                                                                                                                                                                                                                                                                                                                                                                                                                                                                                                                                                                                                                                                                                                                                                                                                                                                                                                                                                                                                                                                                                                                                                                                                                                                                                                                                                                                                                                                                                                                                                                                                                                                                                                                                                                                                                                                                                                                                                                                                                                                                                                                                                                                                                                                                                                                                                                                                                                                                                                                                                                                                                                                                                                                                                                                                                                                                                                                                                                                                                                                                                                                                                                                                                                                                                                                                                                                                                                                                                                                                                                                                                                                                                                                                                                                                                                                                                                                                                                                                                                                                                                                                                                                                                                                                                                                                                                                                                                                                                                                                                                                                                                                                                                                                                                                                                                                                                                                                                                                                                                                                                                                                                                                                                                                                                                                                                                                                                                                                                                                                                                                                                                                                                                                                                                                                                                                                                                                                                                                                                                                                                                                                                                                                                                                                                                                                                                                                                                                                                                                                                                                                                                                                                                                                                                                                                                                                                                                                                                                                                                                                                                                                                                                                                                                                                                                                                                                                                                                                                                                                                                                                                                                                                                                                                                                                                                                                                                                                                                                                                                                                                                                                                                                                                                                                                                                                                                                                                                                                                                                                                                                                                                                                                                                                                                                                                                                                                                                                                                                                                                                                                                                                                                                                                                                                                                                                                                                                                                                                                                                                                                                                                                                                                                                                                                                                                                                                                                                                                                                                                                                                                                                                                                                                                                                                                                                                                                                                                                                                                                    
                                                                                                                                                                                                                                                                                                                                                                                                                                                                                                                                                                                                                                                                                                                                                                                                                                                                                                                                                                                                                                                                                                                                                                                                                                                                                                                                                                                                                                                                                                                                                                                                                                                                                                                                                                                                                                                                                                                                                                                                                                                                                                                                                                                                                                                                                                                                                                                                                                                                                                                                                                                                                                                                                                                                                                                                                                                                                                                                                                                                                                                                                                                                                                                                                                                                                                                                                                                                                                                                                                                                                                                                                                                                                                                                                                                                                                                                                                                                                                                                                                                                                                                                                                                                                                              \subsection{Case 9: $N_e>N$, $N_e<N_c\leq N+N_e$, and $N+N_c-N_e$ is even}{\label{AchScheme9}} In Sections \ref{AchScheme7} and \ref{AchScheme8}, we observe that the flat s.d.o.f. range extends to $N_c=N_e$, and not $N_c=N$ as in Sections \ref{AchScheme2} and \ref{AchScheme3}.  Achieving the alignment of information and cooperative jamming at the eavesdropper requires that $N_c>N_e$ in order for the cooperative jammer to begin sending some jamming signals invisible to the legitimate receiver. 
                                                                                                                                                                                                                                                                                                                                                                                                                                                                                                                                                                                                                                                                                                                                                                                                                                                                                                                                                                                                                                                                                                                                                                                                                                                                                                                                                                                                                                                                                                                                                                                                                                                                                                                                                                                                                                                                                                                                                                                                                                                                                                                                                                                                                                                                                                                                                                                                                                                                                                                                                                                                                                                                                                                                                                                                                                                                                                                                                                                                                                                                                                                                                                                                                                                                                                                                                                                                                                                                                                                                                                                                                                                                                                                                                                                                                                                                                                                                                                                                                                                                                                                                                                                                                                              For this case, in addition to choosing its precoding matrix jointly with the transmitter to satisfy the alignment condition, the cooperative jammer chooses its precoder to send $N_c-N_e$ jamming streams invisible to the receiver. The s.d.o.f. here is integer valued, for which we utilize Gaussian streams. 
                                                                                                                                                                                                                                                                                                                                                                                                                                                                                                                                                                                                                                                                                                                                                                                                                                                                                                                                                                                                                                                                                                                                                                                                                                                                                                                                                                                                                                                                                                                                                                                                                                                                                                                                                                                                                                                                                                                                                                                                                                                                                                                                                                                                                                                                                                                                                                                                                                                                                                                                                                                                                                                                                                                                                                                                                                                                                                                                                                                                                                                                                                                                                                                                                                                                                                                                                                                                                                                                                                                                                                                                                                                                                                                                                                                                                                                                                                                                                                                                                                                                                                                                                                                                                                                        
The transmitted signals are given by (\ref{eq:Xt1_Xc1}) with $d=l=\frac{N+N_c-N_e}{2}$, and $\Ut,\Vc$ are defined as in Section \ref{AchScheme6}. Let $\Pt=\left[\Ptone\;\Pttwo\right]$, and $\Pc=\left[\Pcone\;\Pctwo\right]$, where $\Ptone\in\mathbb{C}^{N\times g}$, $\Pttwo\in\mathbb{C}^{N\times (N_c-N_e)}$, $\Pcone\in\mathbb{C}^{N_c\times g}$, $\Pctwo\in\mathbb{C}^{N_c\times (N_c-N_e)}$, and $g=\frac{N_e+N-N_c}{2}$. The matrices $\Pt$ and $\Pc$ are chosen as follows. Let $\bold{G}=\left[\Gt\;-\Gc\right]\in\mathbb{C}^{N_e\times (N+N_c)}$, and let 
${\bold{G}}'\in\mathbb{C}^{(N_e+N)\times (N+N_c)}$ be expressed as  
\begin{align}
\label{eq:Ach_9_1}
{\bold{G}}'=\begin{bmatrix}\Gt&-\Gc\\\;\bz_{N\times N}&\Hc\\\end{bmatrix}.
\end{align}
Let $\bold{Q}'\in\mathbb{C}^{(N+N_c)\times (N_c-N_e)}$ be randomly chosen such that its columns span $\mathcal{N}(\bold{G}')$, and let the columns of the matrix $\bold{Q}\in\mathbb{C}^{(N+N_c)\times g}$ be randomly chosen as linearly independent vectors in $\mathcal{N}(\bold{G})$, and not in $\mathcal{N}(\bold{G}')$. Write the matrix $\bold{Q}$ as $\bold{Q}=\left[\bold{Q}_1^T\;\bold{Q}_2^T\right]^T$, and the matrix $\bold{Q}'$ as $\bold{Q}'=\left[\bold{Q}_1'^T\;\bold{Q}_2'^T\right]^T$, where $\bold{Q}_1\in\mathbb{C}^{N\times g}$, $\bold{Q}_2\in\mathbb{C}^{N_c\times g}$, $\bold{Q}'_1\in\mathbb{C}^{N\times (N_c-N_e)}$, and $\bold{Q}'_2\in\mathbb{C}^{N_c\times (N_c-N_e)}$. Set $\Ptone=\bold{Q}_1$, $\Pttwo=\bold{Q}'_1$, $\Pcone=\bold{Q}_2$, and $\Pctwo=\bold{Q}'_2$. 

This choice of $\Pt$ and $\Pc$ results in $\Gt\Pt=\Gc\Pc$ and $\Hc\Pctwo=\bz_{N\times (N_c-N_e)}$. Thus, the received signals at the receiver and eavesdropper are given by
\begin{align}
\label{eq:Yr_9_1}
\Yr&=\begin{bmatrix}\Ht\Pt&\Hc\Pcone\\\end{bmatrix}\begin{bmatrix}\Ut\\\Vconeg\\\end{bmatrix}+\Zr\\
\label{eq:Ye_9_1}
\Ye&=\Gc\Pc(\Ut+\Vc)+\Ze.
\end{align}
Using (\ref{eq:Ye_9_1}), and similar to going from (\ref{eq:Ach_1_3}) to (\ref{eq:Ach_1_8}), we have 
\begin{align}
\label{eq:Ach_9_2}
I(\Xt;\Ye)\leq \frac{N+N_c-N_e}{2}.
\end{align}

Because of the assumption of randomly generated channel gains, each of $\Pt$ and $\Pc$ is full column rank a.s. Using Lemma \ref{lemma1}, we have the matrix $\left[\Ht\Pt\;\;\Hc\Pcone\right]$ is full column rank a.s., and hence, using (\ref{eq:Yr_9_1}), we have 
\begin{align}
\label{eq:Ach_9_3}
I(\Xt;\Yr)\geq \frac{N+N_c-N_e}{2}\log P+o(\log P).
\end{align}
Thus, using (\ref{eq:Ach_9_2}), (\ref{eq:Ach_9_3}), (\ref{eq:AchSecRate}), and (\ref{eq:sdof}), the s.d.o.f. is lower bounded as $D_s\geq \frac{N+N_c-N_e}{2}$.

\subsection{Case 10: $N_e>N$, $N_e<N_c\leq N+N_e$, and $N+N_c-N_e$ is odd}{\label{AchScheme10}}
The s.d.o.f. for this case is not an integer, and we have $N_c>N_e$, and hence, we utilize here precoding as in Section \ref{AchScheme9}, and signaling and decoding scheme as in Section \ref{AchScheme8}; $\Ut,\Vc$ are defined as in Section \ref{AchScheme8}, and $\Pt,\Pc$ are chosen as in Section \ref{AchScheme9}, with $d=\frac{N+N_c-N_e+1}{2}$ and $g=\frac{N_e+N-N_c+1}{2}$. Using the same decoding scheme as in Section \ref{AchScheme8}, we obtain that the s.d.o.f. is lower bounded as $D_s\geq \frac{N+N_c-N_e}{2}$ for this case, which completes the achievability proof of (\ref{eq:conv_3_2}). Thus, we have completed the proof for Theorem \ref{Thm1}.     

\section{Extending to the General Case: Theorem \ref{Thm2}}\label{Thm2_Proof}
\subsection{Converse}\label{Thm2_Proof_Conv}
The converse proof for Theorem \ref{Thm2} follows the same steps as in Section \ref{Conv_Proof}. In particular, we derive the following two upper bounds which hold for two different ranges of $N_c$.
\subsubsection{$0\leq N_c\leq N_e$}
Similar to Section \ref{Conv_Proof_1}, we have 
\begin{align}
\label{eq:thm1proof_1}
R_s\leq C_s(P)=\rho\log P+o(\log P),
\end{align}
where, for $0\leq N_c\leq \left[N_e-[N_t-N_r]^+\right]^+$, $\rho=[N_c+N_t-N_e]^+$. Since $[N_c+N_t-N_e]^+\leq N_r$ for $\left[N_e-[N_t-N_r]^+\right]^+\leq N_c\leq N_e$, we have, for $0\leq N_c\leq N_e$,
\begin{align}
\label{eq:thm1proof_2}
D_s \leq \min\{N_r,[N_c+N_t-N_e]^+\}.
\end{align}

\subsubsection{$N_r+[N_e-N_t]^+ < N_c\leq 2\min\{N_t,N_r\}+N_e-N_t$}
Following the same steps as in Section \ref{Conv_Proof_2}, where the two cases we consider here are $N_e\leq N_t$ and $N_e>N_t$, the s.d.o.f. for this range of $N_c$ is upper bounded as 
\begin{align}
\label{eq:thm1proof_3}
D_s\leq\frac{N_c+N_t-N_e}{2}.
\end{align}
Note that, when $N_e>N_t$, this bound holds for $N_c>N_r+N_e-N_t$ so that the number of antennas at the cooperative jammer in the modified channel, c.f. (\ref{eq:modified_channel}), is larger than $N_r$, i.e., $N_c+N_t-N_e>N_r$.

\subsubsection{Obtaining the upper bound}\label{Obtain_bound_thm2}
For each of the following cases, we use the two bounds in (\ref{eq:thm1proof_2}) and (\ref{eq:thm1proof_3}) to obtain the upper bound for the s.d.o.f. 
\begin{enumerate}[i)]
\item $N_t\geq N_r+N_e$\\
For this case, we use the trivial bound for the s.d.o.f., $D_s\leq N_r$ for all the values of $N_c$.

\item $N_r\geq N_t\geq N_e$ and $N_r\geq N_t+N_e$ \\
Using the bound in (\ref{eq:thm1proof_2}), we have
\begin{align*}
D_s\leq N_c+N_t-N_e, {\text{ for }}0\leq N_c\leq N_e,
\end{align*}
where at $N_c=N_e$, we have $D_s\leq N_t$, which is the maximum achievable s.d.o.f. for this case. 

\item $N_t\geq N_e$ and $N_t-N_e<N_r< N_t+N_e$\\
Combining the bounds in (\ref{eq:thm1proof_2}) and (\ref{eq:thm1proof_3}), as in Section \ref{Conv_Proof_3}, yields
\begin{align}
D_s\leq \begin{cases}
N_c+N_t-N_e,\;\;\; 0\leq N_c\leq \frac{N_r+N_e-N_t}{2}\\
\frac{N_r+N_t-N_e}{2},\;\;\;\frac{N_r+N_e-N_t}{2}\leq N_c\leq N_r\\
\frac{N_c+N_t-N_e}{2},\;\;\; N_r\leq N_c\leq 2\min\{N_t,N_r\}+N_e-N_t.
\end{cases}
\end{align}

\item $N_e> N_t$ and $N_r\geq 2N_t$ \\
Using the bound in (\ref{eq:thm1proof_2}), we have
\begin{align*}
D_s\leq [N_c+N_t-N_e]^+, {\text{ for }}0\leq N_c\leq N_e.
\end{align*}

\item $N_e> N_t$ and $N_r<2N_t$\\
By combining the bounds in (\ref{eq:thm1proof_2}) and (\ref{eq:thm1proof_3}), we have 
\begin{align}
D_s\leq \begin{cases}
[N_c+N_t-N_e]^+,\;\;\; 0\leq N_c\leq \frac{N_r}{2}+N_e-N_t\\
\frac{N_r}{2},\quad\;\;\;\frac{N_r}{2}+N_e-N_t\leq N_c\leq N_r+N_e-N_t\\
\frac{N_c+N_t-N_e}{2},\;\;\; N_r+N_e-N_t\leq N_c\leq 2\min\{N_t,N_r\}+N_e-N_t.
\end{cases}
\end{align}
\end{enumerate}

One can easily verify that the cases cited above cover all possible combinations of number of antennas at various terminals. By merging the upper bounds for these cases in one expression, we obtain (\ref{eq:thm2}) as the upper bound for the s.d.o.f. of the channel. 

\subsection{Achievability}\label{Thm2_Proof_Ach} 
The s.d.o.f. for the channel when $N_t$ is not equal to $N_r$, given in (\ref{eq:thm2}), is achieved using techniques similar to what we presented in Section \ref{AchSchemes}. There are few cases, of the number of antennas, where the achievability is straight forward. One such case is when $N_t\geq N_r+N_e$, where the transmitter can send $N_r$ Gaussian information streams invisible to the eavesdropper, and the maximum possible s.d.o.f. of the channel, i.e., $N_r$, is achieved without the help of the cooperative jammer, i.e., $N_c=0$. Another case is when $N_r\geq N_t+\min\{N_t,N_e\}$, where the 
receiver's signal space is sufficient for decoding the information and jamming streams, at high SNR, for all $0\leq N_c\leq N_e$, arriving at the s.d.o.f. of $N_t$ (the maximum possible s.d.o.f.) at $N_c=N_e$. Thus, there is no constant period in the s.d.o.f. characterization for this case where the s.d.o.f. keeps increasing by increasing $N_c$, and Gaussian signaling and cooperative jamming are sufficient to achieve the s.d.o.f. of the channel.  

We consider the five cases of the number of antennas at the different terminals listed in Section \ref{Obtain_bound_thm2}. In the following, we summarize the achievable schemes for these cases. Let $d$ and $l$ denote the number of information and cooperative jamming streams. $\Pt,\Pc$ are the precoding matrices at the transmitter and  the cooperative jammer. 
 
\begin{enumerate}[i)]
\item $N_t\geq N_r+N_e$\\
The transmitter sends $N_r$ Gaussian information streams over $\mathcal{N}(\Gt)$. $D_s=N_r$ is achievable at $N_c=0$.

\item $N_r\geq N_t\geq N_e$ and $N_r\geq N_t+N_e$\\
For $0\leq N_c\leq N_e$, $d=N_c+N_t-N_e$ and $l=N_c$ Gaussian streams are transmitted. Choose $\Pt$ to send $N_t-N_e$ information streams over $\mathcal{N}(\Gt)$ and align the remaining information streams over cooperative jamming streams at the eavesdropper. $D_s=N_c+N_t-N_e$.

\item $N_t\geq N_e$ and $N_t-N_e<N_r< N_t+N_e$:
\begin{enumerate}[1)]
\item For $0\leq N_c\leq \frac{N_r+N_e-N_t}{2}$:\\
The same scheme as in case (ii) is utilized. $D_s=N_c+N_t-N_e$.

\item For $\frac{N_r+N_e-N_t}{2}< N_c\leq N_r$ and $N_r+N_t-N_e$ is even:\\ The same scheme as in case (iii-1), with $d=\frac{N_r+N_t-N_e }{2}$ and $l=\frac{N_r+N_e-N_t}{2}$, is utilized. The cooperative jammer uses only $\frac{N_r+N_e-N_t}{2}$ of its $N_c$ antennas. $D_s=\frac{N_r+N_t-N_e}{2}$.

\item For $\frac{N_r+N_e-N_t}{2}<N_c\leq N_r$ and $N_r+N_t-N_e$ is odd:\\
$d=\frac{N_r+N_t-N_e+1}{2}$ and $l=\frac{N_r+N_e-N_t+1}{2}$ structured streams, as defined in Section \ref{AchScheme3}, are transmitted. The cooperative jammer uses only $\frac{N_r+N_e-N_t+1}{2}$ of its $N_c$ antennas. $\Pt$ is chosen as in case (ii). The legitimate receiver uses the projection and cancellation technique, as in Section \ref{AchScheme3}. $D_s=\frac{N_r+N_t-N_e}{2}$.

\item For $N_r< N_c\leq 2\min\{N_t,N_r\}+N_e-N_t$ and $N_c+N_t-N_e$ is even:\\
$d=\frac{N_c+N_t-N_e}{2}$ and $l=\frac{N_c+N_e-N_t}{2}$ Gaussian streams are transmitted. The cooperative jammer chooses $\Pc$ to send $N_c-N_r$ cooperative jamming streams over $\mathcal{N}(\Hc)$. $\Pt$ is chosen as in case (ii). $D_s=\frac{N_c+N_t-N_e}{2}$.

\item For $N_r< N_c\leq 2\min\{N_t,N_r\}+N_e-N_t$ and $N_c+N_t-N_e$ is odd:\\
$d=\frac{N_c+N_t-N_e+1}{2}$ and $l=\frac{N_c+N_e-N_t+1}{2}$ structured streams are transmitted. $\Pt,\Pc$ are chosen as in case (iii-4). The legitimate receiver uses the projection and cancellation technique. $D_s=\frac{N_c+N_t-N_e}{2}$.
\end{enumerate}

\item $N_e> N_t$ and $N_r\geq 2N_t$ \\
For $0\leq N_c\leq N_e$, $d=l=[N_c+N_t-N_e]^+$ Gaussian streams are transmitted. Both $\Pt,\Pc$ are chosen to align the information streams over the cooperative jamming streams at the eavesdropper as in Section \ref{AchScheme6}. $D_s=[N_c+N_t-N_e]^+$.

\item $N_e>N_t$ and $N_r<2N_t$:
\begin{enumerate}[1)]
\item For $0\leq N_c\leq \frac{N_r}{2}+N_e-N_t$:\\
The same scheme as in case (iv) is utilized. $D_s=[N_c+N_t-N_e]^+$.
 
\item For $\frac{N_r}{2}+N_e-N_t< N_c \leq N_r+N_e-N_t$ and $N_r$ is even:\\
$d=l=\frac{N_r}{2}$ Gaussian streams are transmitted. $\Pt,\Pc$ are chosen as in case (iv). $D_s=\frac{N_r}{2}$.
\item For $\frac{N_r}{2}+N_e-N_t< N_c \leq N_r+N_e-N_t$ and $N_r$ is odd:\\
$d=l=\frac{N_r+1}{2}$ structured streams are transmitted. $\Pt,\Pc$ are as in case (iv). The legitimate receiver uses the projection and cancellation technique. $D_s=\frac{N_r}{2}$.

\item For $N_r+N_e-N_t< N_c\leq 2\min\{N_t,N_r\}+N_e-N_t$ and $N_c+N_t-N_e$ is even:\\
$d=l=\frac{N_c+N_t-N_e}{2}$ Gaussian streams are transmitted. Both $\Pt,\Pc$ are chosen to align the information and the cooperative jamming streams at the eavesdropper. $\Pc$ is also chosen to send $N_c-N_r$ cooperative jamming streams over $\mathcal{N}(\Hc)$ as in Section \ref{AchScheme9}. $N_c>N_r+N_e-N_t$ achieves the above two conditions. $D_s=\frac{N_c+N_t-N_e}{2}$. 
    
\item For $N_r+N_e-N_t< N_c\leq 2\min\{N_t,N_r\}+N_e-N_t$ and $N_c+N_t-N_e$ is odd:\\
$d=l=\frac{N_c+N_t-N_e+1}{2}$ structured streams are transmitted. $\Pt,\Pc$ are chosen as in case (v-4). The receiver uses the projection and cancellation technique. $D_s=\frac{N_c+N_t-N_e}{2}$.
\end{enumerate}
\end{enumerate}

Using the achievable schemes described above for the different cases of the number of antennas, and their analysis as in Section \ref{AchSchemes}, we have (\ref{eq:thm2}) as the achievable s.d.o.f., which completes the proof for theorem \ref{Thm2}. 

\section{Discussion}\label{Discussion} 
At this point, it is useful to discuss the results and the implications of this work. Theorem \ref{Thm1}, c.f. (\ref{eq:thm1}), shows the behavior of the s.d.o.f., for an $(N\times N\times N_e) $ multi-antenna Gaussian wire-tap channel with an $N_c$-antenna cooperative jammer, associated with increasing $N_c$ form $0$ to $N+N_e$. The s.d.o.f. first increases linearly by increasing $N_c$ from $0$ to $N_e-\lceil\frac{\min\{N,N_e\}}{2}\rceil$, that is to say adding one antenna at the cooperative jammer provided the system to have one additional degrees of freedom. The s.d.o.f. remains constant in the $N_c$ range of $N_e-\lfloor\frac{\min\{N,N_e\}}{2}\rfloor$ to $\max\{N,N_e\}$, and starts to increase again for $N_c$ from $\max\{N,N_e\}$ to $N+N_e$, until the s.d.o.f. arrives at its maximum value, $N$, at $N_c=N+N_e$. This behavior transpires both when the eavesdropper antennas are fewer or more than that of the legitimate receiver.

The reason for the flat s.d.o.f. range is as follows: At high SNR, achieving the secrecy constraint requires i) the entropy of the cooperative jamming signal, $\Xcn$, to be greater than or equal to that of the information signal visible to the eavesdropper, and ii) $\Xcn$ to completely cover the information signal, $\Xtn$, at the eavesdropper. For $N_e\leq N$, part of $\Xtn$ can be sent invisible to the eavesdropper, and the information signal visible to the eavesdropper can be covered by jamming for all $N_c$. For $0\leq N_c\leq \frac{N_e}{2}$, the spatial resources at the receiver are sufficient, at high SNR, for decoding information and jamming signals which satisfy the above constraints. Thus, increasing the possible entropy of $\Xcn$ by increasing $N_c$ from $0$ to $\left\lfloor \frac{N_e}{2}\right\rfloor$ allows for increasing the entropy of $\Xtn$, and hence, the achievable secrecy rate and the s.d.o.f. increase. At $N_c=\left\lceil\frac{N_e}{2}\right\rceil$, the possible entropy of $\Xcn$ and, correspondingly, the maximum possible entropy of $\Xtn$, result in information and jamming signals which completely occupy the receiver's signal space. Thus, increasing the possible uncertainty of $\Xcn$ by increasing $N_c$ from $\left\lceil\frac{N_e}{2}\right\rceil$ to $N$ is useless, since, in this range, $\Xcn$ is totally observed by the receiver which has its signal space already full at $N_c=\left\lceil\frac{N_e}{2}\right\rceil$. 

Increasing $N_c$ over $N$ increases the possible entropy of $\Xcn$ and simultaneously increases the part of $\Xcn$ that can be transmitted invisible to the receiver, leaving more space for $\Xtn$ at the receiver. This allows for increasing the secrecy rate, and hence, the s.d.o.f. starts to increase again. For $N_e>N$, the s.d.o.f. is equal to zero for all $0\leq N_c\leq N_e-N$, where $\Xcn$ can not cover the information at the eavesdropper for this case. The s.d.o.f. starts to increase again, after the flat range, at $N_c>N_e$, since sending jamming signals invisible to the receiver while satisfying the covering condition at the eavesdropper requires that $N_c>N_e$. 

The difference in the slope for the increase in the s.d.o.f. in the ranges before and after the flat range, for both $N_e\leq N$ and $N_e>N$, can be explained as follows. For $0\leq N_c\leq N_e-\frac{\min\{N,N_e\}}{2}$, each additional antenna at the cooperative jammer allows for utilizing two more spatial dimensions at the receiver;  one spatial dimension is used for the jamming signal and the other is used for the information signal. By contrast, for $\max\{N,N_e\}< N_c\leq N+N_e$, each additional antenna at the cooperative jammer sets one spatial dimension at the receiver free from jamming, and this spatial dimension is shared between the extra cooperative jamming and information streams.

It is important to note that the result that suggests that increasing the cooperative jammer antennas is not useful in the range $N_e-\frac{\min\{N,N_e\}}{2}<N_c\leq \max\{N,N_e\}$ applies only to the prelog of the secrecy capacity, i.e., is specific to the high SNR behavior. This should not be taken to mean that additional antennas do not improve secrecy rate, but only the secrecy rate scaling with power in the high SNR. 

Theorem \ref{Thm2} generalizes the results above to the case where the number of transmit antennas at the transmitter, $N_t$, is not equal to the number of receive antennas at the legitimate receiver, $N_r$. Although the maximum possible s.d.o.f. of the channel for this case is limited to $\min\{N_t,N_r\}=N_d$, increasing $N_t$ over $N_r$, or increasing $N_r$ over $N_t$, do change the behavior of the s.d.o.f. associated with increasing $N_c$ until the maximum possible s.d.o.f. is reached. Let us start at $N_t=N_r=N_d$. For $N_t\geq N_e$, increasing $N_t$ over $N_d=N_r$ increases the number of the information streams that can be sent invisible to the eavesdropper, and hence the s.d.o.f. without the help of the CJ, i.e., $N_c=0$, increases. This results in increasing the range of $N_c$ for which the s.d.o.f. remains constant by increasing $N_c$, since the receiver's signal space gets full at a smaller $N_c$ and remains full until $N_c$ is larger than $N_d=N_r$. In addition, increasing $N_t$ over $N_d$, when $N_t\geq N_e$, results in decreasing the value of $N_c$ at which the maximum s.d.o.f. of the channel, $N_d$, is achievable, arriving at $N_t\geq N_r+N_e$, where the s.d.o.f. of $N_d$ is achievable without the help of the CJ. When $N_e>N_t$, increasing $N_t$ over $N_d$ decreases the value of $N_c$ at which the s.d.o.f. is positive, and decreases the value of $N_c$ at which the s.d.o.f. of $N_d$ is achievable, arriving at $N_t>N_e$, where the channel renders itself to the previous case. For both the cases $N_t\geq N_e$ and $N_t<N_e$, increasing $N_r$ over $N_d=N_t$, results in increasing the available space at the receiver's signal space, and hence the constant period decreases, arriving at $N_r\geq N_t+N_e$ when $N_t\geq N_e$, or at $N_r\geq 2N_t$ when $N_e> N_t$, where the constant period vanishes.
 
\section{Conclusion}\label{Con} 
In this paper, we have studied the multi-antenna wire-tap channel with a $N_c$-antenna cooperative jammer, $N_t$-antenna transmitter, $N_r$-antenna receiver, and $N_e$-antenna eavesdropper. We have completely characterized the s.d.o.f. for this channel for all possible values of the number of antennas at the cooperative jammer, $N_c$. We have shown that when the s.d.o.f. of the channel is integer valued, it can be achieved by linear precoding at the transmitter and cooperative jammer, Gaussian signaling both for transmission and jamming, and linear processing at the legitimate receiver. By contrast, when the s.d.o.f. is not an integer, we have shown that a scheme which employs structured signaling both at the transmitter and cooperative jammer, along with joint signal space and signal scale alignment achieves the s.d.o.f. of the channel. We have seen that, when $N_t\geq N_e$, the transmitter uses its precoder to send a part of its information signal invisible to the eavesdropper, and to align the remaining part over jamming at the eavesdropper, while the cooperative jammer uses its precoder to send a part of its jamming signal invisible to the receiver, whenever possible. When $N_e>N_t$, more intricate precoding at the transmitter and cooperative jammer is required, where both the transmitter and cooperative jammer choose their precoders to achieve the alignment of information and jamming at the eavesdropper, and simultaneously, the cooperative jammer designs its precoder, whenever possible, to send a part of the jamming signal invisible to the receiver. The converse was established by allowing for full cooperation between the transmitter and cooperative jammer for a certain range of $N_c$, and by incorporating both the secrecy and reliability constraints, for the other values of $N_c$. We note that while this paper settles the degrees of freedom of this channel, its secrecy capacity is still open. Additionally, while the model considered here assumes channels to be known, universal secrecy as in \cite{he2014mimo} should be considered in the future. 
\section*{Appendix A\\Choice of $\Kt$ and $\Kc$} \label{App:AppendixA}
The covariance matrices $\Kt$ and $\Kc$ are chosen so that they are {\it{positive definite}}, i.e., $\Kt,\Kc\succ \bz$, and hence non-singular, in order to guarantee the finiteness of $h(\Zttilde)$ and $h(\Zctilde)$ in (\ref{eq:conv_2_11}). In addition, positive definite $\Kt$ and $\Kc$ result in positive definite $\bold{\Sigma}_{\Zone}$ and $\bold{\Sigma}_{\Ztwo}$, and hence, $h(\Zone)$ and $h(\Ztwo)$ in (\ref{eq:conv_2_13}) are also finite. 

For $\bI_{N_e}-\Gt\Kt\GtH$ to be a valid covariance matrix for $\Zetilde$ in (\ref{eq:conv_2_15}), $\Kt$ has to satisfy $\Gt\Kt\GtH\preceq \bI_{N_e}$, which is equivalent to
\begin{align}
\label{eq:AppendixA_2}
||\Kt^{\frac{1}{2}}\GtH||\leq 1.
\end{align} 
Recall that $||\Kt^{\frac{1}{2}}\GtH||$ is the induced norm for the matrix $\Kt^{\frac{1}{2}}\GtH$.

Similarly, for $\bI_N-\Hc\Kc\HcH$, $\bI_{N_e}-\Gt\Kt\GtH-\Gc\Kc\GcH$, and $\bI_N-\Hctwo'\Kc'\HctwodH$ to be valid covariance matrices for $\Zrtilde,\Zedtilde$, and $\Zrdtilde$, in (\ref{eq:conv_2_25}), (\ref{eq:conv_2_38}), (\ref{eq:conv_2_44_1}), $\Kt,\Kc,\Kc'$ have to satisfy  
\begin{align} 
\label{eq:AppendixA_3}
||\Kc^{\frac{1}{2}}\HcH||\leq 1,\quad ||\Kt^{\frac{1}{2}}\GtH||^2+||\Kc^{\frac{1}{2}}\GcH||^2&\leq 1,\quad \text{and } ||\Kc'^{\frac{1}{2}}\HctwodH||\leq 1. 
\end{align}
In order to satisfy the conditions (\ref{eq:AppendixA_2}) and (\ref{eq:AppendixA_3}), we choose $\Kt=\rho^2 \bI_{N}$, $\Kc=\rho^2 \bI_{K}$, where
\begin{align}
0<\rho&\leq 1/\max\left\{||\GtH||,||\HcH||,\sqrt{||\GtH||^2+||\GcH||^2},||\HctwodH||\right\}\\
&=1/\max\left\{||\HcH||,\sqrt{||\GtH||^2+||\GcH||^2}\right\}.
\end{align}
 
\section*{Appendix B\\Derivation of (\ref{eq:conv_2_33}), (\ref{eq:conv_2_34}), and (\ref{eq:conv_2_48})} \label{App:AppendixB}
In order to upper bound $h(\Yrk(i))$, for all $i=1,2,\cdots,n$ and $k=1,2,\cdots,N$, we first upper bound the variance of $\Yrk(i)$, denoted by $\var \{\Yrk(i)\}$. Let $\bh_{t,k}^{r}$ and $\bh_{c,k}^{r}$ denote the transpose of the $k$th row vectors of $\Ht$ and $\Hc$, respectively. Let $\Zr(i)=\left[Z_{r,1}(i)\cdots Z_{r,N}(i)\right]^T$. Using (\ref{eq:Yr1}), $\Yrk(i)$ is expressed as
\begin{align}
\label{eq:AppendixB_1}
\Yrk(i)=\bh_{t,k}^{r^T}\Xt(i)+\bh_{c,k}^{r^T}\Xc(i)+Z_{r,k}(i).
\end{align}
Thus, $\var\{\Yrk(i)\}$ can be bounded as 
\begin{align}
\label{eq:AppendixB_2}
\var \left\{\Yrk(i)\right\}&\leq \E\left\{\Yrk(i)\Yrkcon(i)\right\}\\
\label{eq:AppendixB_3}
&=\E\left\{|\bh_{t,k}^{r^T}\Xt(i)|^2\right\}+\E\left\{|\bh_{c,k}^{r^T}\Xc(i)|^2\right\}+\E\left\{|Z_{r,k}(i)|^2\right\}\\
\label{eq:AppendixB_4}
&\leq ||\bh_{t,k}^{r}||^2 \;\E\left\{||\Xt(i)||^2\right\}+||\bh_{c,k}^{r}||^2\;\E\left\{||\Xc(i)||^2\right\}+1\\
\label{eq:AppendixB_5}
&\leq 1+\left(||\bh_{t,k}^{r}||^2+||\bh_{c,k}^{r}||^2\right) P,
\end{align}
where (\ref{eq:AppendixB_4}) follows from Cauchy-Schwarz inequality and monotonicity of expectation, and (\ref{eq:AppendixB_5}) follows from the power constraints at the transmitter and cooperative jammer. 

Define $h^2=\underset{k}{\max}\;\left(||\bh_{t,k}^{r}||^2+||\bh_{c,k}^{r}||^2\right)$. Since $h(Y_{r,k}(i))$ is upper bounded by the entropy of a complex Gaussian random variable with the same variance, we have, for all $i=1,2,\cdots,n$ and $k=1,2,\cdots,N$,
\begin{align}
\label{eq:AppendixB_6}
h(\Yrk(i))&\leq \log 2\pi e \left(1+\left(||\bh_{t,k}^{r}||^2+||\bh_{c,k}^{r}||^2\right) P\right)\\
\label{eq:AppendixB_7}
&\leq \log 2\pi e +\log (1+h^2P).
\end{align}

Similarly, we have 
\begin{align}
\bar{Y}_{r,k}(i)=\bh_{t,k}^{r^T}\Xt(i)+\bh_{c,k}'^{r^T}\Xctwo'(i)+Z_{r,k}(i),
\end{align}
where $\bh_{c,k}'^r$ is the transpose of the $k$-th row vector of $\Hctwo'$. Thus, we have, 
\begin{align}
h(\bar{Y}_{r,k}(i))\leq \log 2\pi e +\log (1+\bar{h}^2 P),
\end{align}
where $\bar{h}^2=\underset{k}{\max}\;\left(||\bh_{t,k}^r||^2+||\bh_{c,k}'^r||^2\right)$.

Next, we upper bound $h(\tilde{X}_{t,k}(i))$. The power constraint at the transmitter, for  $i=1,2,\cdots,n$, is $\E\left\{\XtH(i)\;\Xt(i)\right\}=\sum_{k=1}^N \E\left\{|X_{t,k}(i)|^2\right\}\leq P$. Thus, $\E\left\{|X_{t,k}(i)|^2\right\}\leq P$ for all $i=1,2,\cdots,n$, and $k=1,2,\cdots,N$ . Recall that $\tilde{X}_{t,k}(i)=X_{t,k}(i)+\tilde{Z}_{t,k}(i)$, where $X_{t,k}(i)$ and $\tilde{Z}_{t,k}(i)$ are independent, and the covariance matrix of $\Zttilde$ is $\Kt=\rho^2\bI_{N}$, where $0<\rho\leq \min\left\{\frac{1}{||\HcH||},\frac{1}{\sqrt{||\GtH||^2+||\GcH||^2}}\right\}$. Thus, $\var\{\tilde{X}_{t,k}(i)\}$ is upper bounded as 
\begin{align}
\label{eq:AppendixB_8}
\var\{\tilde{X}_{t,k}(i)\}&=\var\{X_{t,k}(i)\}+\var\{\tilde{Z}_{t,k}(i)\}\\
\label{eq:AppendixB_10}
&\leq\E\left\{|X_{t,k}(i)|^2\right\}+\rho^2\leq P+\rho^2.
\end{align}
Thus, for $i=1,2,\cdots,n$ and $k=1,2,\cdots,N$, we have
\begin{align}
\label{eq:AppendixB_11}
h(\tilde{X}_{t,k}(i))\leq \log 2\pi e+\log (\rho^2+P).
\end{align} 
Similarly, using the power constraint at the cooperative jammer, we have, for $i=1,\cdots,n$ and $j=1,\cdots,K$, 
\begin{align}
\label{eq:AppendixB_12}
h(\tilde{X}_{c,j}(i))\leq \log 2\pi e+\log (\rho^2+P).
\end{align}

\section*{Appendix C\\Proof of Lemma \ref{lemma1}} \label{App:AppendixB}
Consider two matrices $\bold{Q}\in\mathbb{C}^{M\times K}$ and $\bold{W}\in\mathbb{C}^{K\times N}$ such that $\bold{Q}$ is full row-rank and $\bold{W}$ has all of its entries independently drawn from a continuous distribution, where $K>N,M$. Let $L=\min\{N,M\}$. We show that $\bold{Q}\bold{W}$ has a rank $L$ a.s. . The matrices $\bold{Q}$ and $\bold{W}$ can be written as 
\begin{align}
\bold{Q}&=\begin{bmatrix}\bold{q}_1&\bold{q}_2&\cdots&\bold{q}_K\\\end{bmatrix},\\ \bold{W}&=\begin{bmatrix}\bold{w}_1&\bold{w}_2&\cdots&\bold{w}_N\\\end{bmatrix},
\end{align} 
where $\bold{q}_1,\bold{q}_2,\cdots,\bold{q}_{K}$ are the $K$ length-$M$ column vectors of $\bold{Q}$, and $\bold{w}_1,\bold{w}_2,\cdots,\bold{w}_{N}$ are the $N$ length-$K$ column vectors of $\bold{W}$.

Let $w_{j,i}$ denotes the entry in the $j$th row and $i$th column of $\bold{W}$. Let $\bold{Q}\bold{W}=[\bold{s}_1\;\bold{s}_2\;\cdots\;\bold{s}_N]$, where $\bold{s}_i$ is a length-$M$ column vector, $i=1,2,\cdots,N$. When $M\geq N$, $\bold{QW}=[\bold{s}_1\;\bold{s}_2\;\cdots\;\bold{s}_L]$, and when $M<N$, $\{\bold{s}_1,\bold{s}_2,\cdots,\bold{s}_L\}$ are the first $L$ columns of $\bold{QW}$. In order to show that the matrix $\bold{QW}$ has rank $L$, we show that, in either case, $\{\bold{s}_1,\bold{s}_2,\cdots,\bold{s}_L\}$ are a.s.  linearly independent, i.e., 
\begin{align}
\label{eq:LinIndepCond}
\sum_{i=1}^L \lambda_i \bold{s}_i=\bz_{M\times 1}
\end{align}
if and only if $\lambda_i=0$ for all $i=1,2,\cdots,L$.

Each $\bold{s}_i$, for $i=1,2,\cdots,L$, can be viewed as a linear combination of the $K$ columns of $\bold{Q}$ with coefficients that are the entries of the $i$th column of $\bold{W}$, i.e.,
\begin{align}
\bold{s}_i=\sum_{j=1}^{K} w_{j,i}\bold{q}_j.
\label{eq:gi}
\end{align}

Using (\ref{eq:gi}), we can rewrite (\ref{eq:LinIndepCond}) as
\begin{align}
\sum_{j=1}^{K} \varphi_j \bold{q}_j=\bz_{M\times 1}
\label{eq:LinIndepCond1}
\end{align}
where, for $j=1,2,\cdots,K$,
\begin{align}
\varphi_j=\sum_{i=1}^{L} \lambda_i w_{j,i}.
\label{eq:mj}
\end{align}
The $K$ columns of $\bold{Q}$ are linearly dependent since each of them is of length $M$ and $K>M$. Thus, equation (\ref{eq:LinIndepCond1}) has infinitely many solutions for $\left\{\varphi_j\right\}_{j=1}^{K}$. 

Each of these solutions for $\varphi_j$'s constitutes a system of $K$ linear equations $\big\{\varphi_j=\sum_{i=1}^{L} \lambda_i w_{j,i},\\j=1,2,\cdots,K\big\}$. The number of unknowns in this system, i.e. $\lambda$'s, is $L$. Since the number of equations in this system, $K$, is greater than the number of unknowns, $L$, this system has a solution for $\left\{\lambda_i\right\}_{i=1}^{L}$ only if the elements $\{w_{j,i}: j=1,2,\cdots,K,\; \text{and } i=1,2,\cdots,L\}$ are dependent. Since the entries of $\bold{W}$ are all randomly and independently drawn from some continuous distribution, the probability that these entries are dependent is zero. 

Moreover, consider the set with infinite cardinality, where each element in this set is a structured $\bold{W}$ that causes the system of equations in (\ref{eq:mj}) to have a solution for $\{\lambda_i\}_{i=1}^L$ for one of the infinitely many solutions of $\{\varphi_j\}_{j=1}^K$ to (\ref{eq:LinIndepCond1}). This set with infinite cardinality has a measure zero in the space $\mathbb{C}^{K\times L}$, since this set is a subspace of $\mathbb{C}^{K\times L}$ with a dimension strictly less than $K\times L$. We conclude that (\ref{eq:LinIndepCond}) a.s.  has no non-zero solution for $\{\lambda_i\}_{i=1}^L$. Thus, $\bold{Q}\bold{W}$ has rank $L$ a.s. 

If $\bold{Q}\bold{W}$ has rank $L$ a.s. , then so does $(\bold{Q}\bold{W})^T=\bold{W}^T\bold{Q}^T$. Setting $\bold{E}_1=\bold{W}^T$ and $\bold{E}_2=\bold{Q}^T$, we have $\bold{E}_1\in\mathbb{C}^{N\times K}$ has all of its entries independently drawn from some continuous distribution, $\bold{E}_2\in\mathbb{C}^{K\times M}$  is full column-rank, $K>N,M$, and $\bold{E}_1\bold{E}_2$ has rank $L=\min\{N,M\}$ a.s. Thus, Lemma \ref{lemma1} is proved. 

\section*{Appendix D\\Derivation of (\ref{eq:Q}) and (\ref{eq:a})} \label{App:AppendixC}
The power constraints at the transmitter and cooperative jammer are $\E\left\{\XtH\Xt\right\}\leq P$ and $\E\left\{\XcH\Xc\right\}\leq P$. Using (\ref{eq:Xt1_Xc1}), we have 
\begin{align}
\label{eq:AppendixD_1}
\E\left\{\XtH\Xt\right\}&=\E\left\{\UtH\PtH\Pt\Ut\right\}\\
\label{eq:AppendixD_3}
&=\sum_{i=1}^{d}\sum_{j=1}^{d}\bp_{t,j}^H\bp_{t,i}\E\left\{U_j^{*}U_i\right\}\\
\label{eq:AppendixD_4}
&=\sum_{i=1}^{d}||\bp_{t,i}||^2\E\left\{|U_i|^2\right\}\\
\label{eq:AppendixD_5}
&=||\bp_{t,1}||^2\E\left\{|U_1|^2\right\}+\sum_{i=2}^{d}||\bp_{t,i}||^2 \left(\E\left\{\Uire^2\right\}+\E\left\{\Uiim^2\right\}\right)\\
\label{eq:AppendixD_6}
&\leq \left(||\bp_{t,1}||^2+2\sum_{i=2}^{d}||\bp_{t,i}||^2\right) a^2 Q^2,
\end{align}
where (\ref{eq:AppendixD_4}) follows since $U_i$ and $U_j$, for $i\neq j$, are independent, and (\ref{eq:AppendixD_6}) follows since $\E\left\{U_1^2\right\}, E\left\{\Uire^2\right\}, E\left\{\Uiim^2\right\}\leq a^2 Q^2$, for $i=2,3,\cdots,d$. 
  
Similarly, using (\ref{eq:Xt1_Xc1}) and (\ref{eq:Ach_2_1}), we have 
\begin{align}
\label{eq:AppendixD_7}
\E\left\{\XcH\Xc\right\}&=\E\left\{\VcH\PcH\Pc\Vc\right\}=\sum_{i=1}^{l}\E\left\{|V_i|^2\right\}\\
\label{eq:AppendixD_9}
&=\E\left\{V_1^2\right\}+\sum_{i=2}^{l}\left(\E\left\{\Vire^2\right\}+\E\left\{\Viim^2\right\}\right)\\
\label{eq:AppendixD_10}
&\leq (2l-1) a^2Q^2.  
\end{align}

From (\ref{eq:AppendixD_6}) and (\ref{eq:AppendixD_10}), in order to satisfy the power constraints, we need that 
\begin{align}
\label{eq:AppendixD_11}
a^2Q^2\leq \gamma^2 P,
\end{align}
where, 
\begin{align}
\label{eq:AppendixD_12}
\gamma^2=\frac{1}{\max\left\{2l-1,||\bp_{t,1}||^2+2\sum_{i=2}^{d}||\bp_{t,i}||^2\right\}}.
\end{align}
Let us choose the integer $Q$ as 
\begin{align}
\label{eq:AppendixD_13}
Q=\left\lfloor P^{\frac{1-\epsilon}{2+\epsilon}}\right\rfloor =P^{\frac{1-\epsilon}{2+\epsilon}}-\nu,
\end{align}
where $\nu$ is a constant which does not depend on the power $P$. Thus,
\begin{align}
\label{eq:AppendixD_14}
a=\gamma P^{\frac{3\epsilon}{2(2+\epsilon)}},
\end{align}
satisfies the condition in (\ref{eq:AppendixD_11}). Thus, the power constraints at the transmitter and cooperative jammer are satisfied.

\bibliographystyle{IEEEtran}
\bibliography{MyLib}

\end{document}